\documentclass[12pt]{article}
\usepackage[round]{natbib}
\usepackage[utf8]{inputenc}
\usepackage{graphicx,amsmath,amsfonts,amssymb,fullpage,url}
\usepackage{algorithm}
\usepackage{algorithmic}

\usepackage{array}
\usepackage{tabularx}
\usepackage{caption}
\usepackage{adjustbox}
\usepackage{enumitem}
\usepackage{bookmark}
\usepackage{hyperref}

\usepackage{tcolorbox}
\usepackage{fancyhdr}

\usepackage{parskip} 
\setlist[itemize]{noitemsep, before={\setlength{\parskip}{0pt}}}
\usepackage{cmbright}
\usepackage[OT1]{fontenc}

\usepackage{xcolor}
\definecolor{myred}{HTML}{E06666}
\definecolor{myblue}{HTML}{299DD3}
\definecolor{mygreen}{HTML}{159B27}
\definecolor{lightsteel}{HTML}{8f8f8c}
\definecolor{mediumsteel}{RGB}{82, 83, 94}
\definecolor{darksteel}{RGB}{54, 69, 79}
\definecolor{darkgreen}{RGB}{0,77,64}
\definecolor{prettycyan}{RGB}{0, 180, 180}  
\definecolor{olive}{HTML}{9CB380}     
\definecolor{orange}{HTML}{FECDAA}     
\definecolor{salmon}{HTML}{FFA3A5}     
\definecolor{andrewpink}{HTML}{DD34A5}
\definecolor{wilkacolor}{HTML}{159B27}

\newcommand{\paperarg}[1]{\textbf{\color{lightsteel} #1}}

\hypersetup{
    colorlinks=true,
    linkcolor=mygreen,  
    urlcolor=salmon,     
    citecolor=prettycyan,  
    filecolor=darkblue     
}

\newtcolorbox{definitionbox}[1]{
  colback=lightsteel,
  colframe=darksteel,
  fonttitle=\bfseries\color{white},
  coltitle=white,
  title=#1,
  sharp corners,
  boxrule=1pt,
  toptitle=1mm,
  bottomtitle=1mm,
  top=3mm,
}

\long\def\andrew#1{{#1}}
\long\def\wilka#1{{#1}}

\title{
\vspace{5em}
    \textbf{Naturalistic Computational Cognitive Science} \\[0.5em]
    {\Large Towards generalizable models and theories \\
     that capture the full range of natural behavior}
}
\author{
    \textbf{Wilka Carvalho}$^\ast$ \\
    Kempner Institute for the Study of Natural and Artificial Intelligence \\
    Harvard University \\
    Cambridge, MA, USA \\
    \texttt{wcarvalho@g.harvard.edu} \\
    \\
    \textbf{Andrew Lampinen}$^\ast$ \\
    Google DeepMind \\
    San Francisco, CA, USA\\
    \texttt{lampinen@google.com} \\
    \\
    $^\ast$equal contribution
}

\date{}
\begin{document}

\maketitle

\clearpage

\begin{abstract}
How can cognitive science build generalizable theories that span the full scope of natural situations and behaviors?
We argue that progress in Artificial Intelligence (AI) offers timely opportunities for cognitive science to embrace experiments with increasingly naturalistic stimuli, tasks, and behaviors; and computational models that can accommodate these changes. We first review a growing body of research spanning neuroscience, cognitive science, and AI that suggests that incorporating a broader range of naturalistic experimental paradigms, and models that accommodate them, may be necessary to resolve some aspects of natural intelligence and ensure that our theories generalize. We review cases from cognitive science and neuroscience where naturalistic paradigms elicit distinct behaviors or engage different processes. We then discuss recent progress in AI that shows that learning from naturalistic data yields qualitatively different patterns of behavior and generalization, and examine how these findings impact the conclusions we draw from cognitive modeling, and can help yield new hypotheses for the roots of cognitive and neural phenomena.
We then suggest that integrating recent progress in AI and cognitive science will enable us to engage with more naturalistic phenomena without giving up experimental control or the pursuit of theoretically grounded understanding. We offer practical guidance on how methodological practices can contribute to cumulative progress in naturalistic computational cognitive science, and illustrate a path towards building computational models that solve the real problems of natural cognition, together with a reductive understanding of the processes and principles by which they do so.
\end{abstract}

\clearpage

\section{Introduction}

Cognitive scientists build models to make our theories concrete---which offers testable predictions, eliminates ambiguities \citep{guest2021computational}, and can reveal unexpected properties of cognition~\citep{mcclelland2009place}. Cognitive models can range from simplified models of a high-level process \citep[e.g.][]{frank2012predicting,cohen1990control}, to task-performing models of a particular domain \citep{newell2012you}, or beyond \citep{newell1994unified}. Many cognitive paradigms---connectionism~\citep{mcclelland1986appeal}, Bayesian inference~\citep{tenenbaum2001generalization,tenenbaum2011grow}, cognitive architectures~\citep[e.g.][]{ritter2019act,laird2019soar}---have sought generalizable models that can explain a broad range of behavior from a simpler set of principles.
However, despite progress, there is still substantial debate on the generalizability of our theories \citep{yarkoni2022generalizability,eckstein2021reinforcement}, whether our models and theories are adequately constrained \citep[e.g.][]{jones2011bayesian,rahnev2018suboptimality}, and calls to accommodate a broader scope of naturalistic phenomena \citep{nastase2020keep,wise2023naturalistic,cisek2024toward}. Our goal in this paper is to motivate and outline a path that we believe will begin to address many of these challenges.

The path we propose is \andrew{inspired} by recent practical and conceptual developments in AI. 
We now have vision models that can classify and segment real-world images without external supervision~\citep{he2020momentum,caron2021emerging}, reinforcement learning models that can learn new tasks in a domain with human-like efficiency~\citep{team2023human}, and language models that can solve many language-specified tasks~\citep{radford2019language,wei2022emergent}. 
\wilka{Surprisingly, such models produce internal representations that capture many features of brain representations \citep{yamins2014performance,schrimpf2021neural}. This has prompted debate on whether such complex models can be explanatory \citep{cao2024explanatorypt1,kanwisher2023using}, what they imply for our theories \citep{hasson2020direct,perconti2020deep,piantadosi2023modern}, and a broader question of how we should test and interpret our theories generally \citep{bowers2023deep,dicarlo2023let,dentella2024testing}. Thus, whether and how this progress can serve cognitive science remains open.
}

\begin{figure}[htp]
    \begin{center}
        \includegraphics[width=\textwidth]{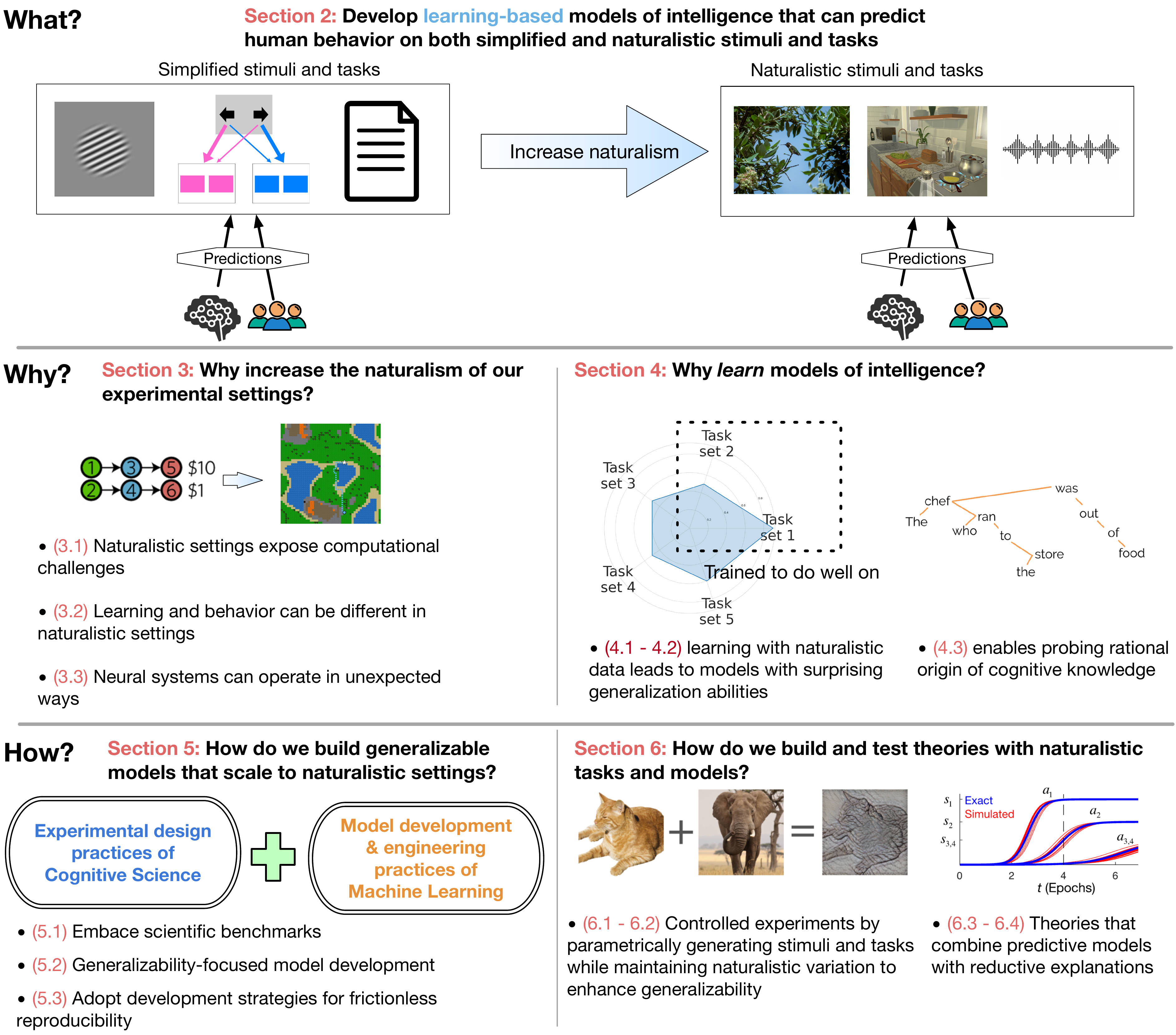} 
        \caption{\textbf{Naturalistic computational cognitive science}: the what, why, and the how. The first section (\S\ref{sec:definitions}) provides an overview of ``naturalistic computational cognitive science''. In (\S\ref{sec:data-benefits}-\S\ref{sec:learning-benefits}), we motivate using naturalistic experimental paradigms and learning-based approaches for cognitive research. The remaining sections (\S\ref{sec:building}-\S\ref{sec:theory}) focus on \textbf{how} to achieve these goals; how to develop models for naturalistic settings, and how to use naturalistic experiments and models as part of explanatory cognitive theories. (Figures reproduced from \citealp{saxe2019mathematical,geirhos2018imagenet,doshi2023cortical}.)}
        \label{fig:overview}
    \end{center}
\end{figure}

In this work, we attempt to weave a thread of arguments that synthesizes these literatures, ranging from cognitive science's motivation and theory development to the practicalities of model engineering and reproducibility.
We articulate a research strategy that we believe will be helpful in enabling theory-driven cognitive science to achieve a deeper understanding of the full range of natural intelligence.
We present an overview of \andrew{the paper and} our arguments in Figure~\ref{fig:overview}.

\wilka{Our paper proceeds as follows. In \S\ref{sec:definitions}, we specify what naturalistic computational cognitive science means in practice. We argue that researchers should broaden their paradigms beyond the settings in which their theoretical constructs are typically tested---by both adding variables that interact with those constructs and by sampling more broadly within existing parameters. Likewise, we should build models that more faithfully approximate the computational task that the natural system is solving.
}

\begin{figure}[htp]
    \begin{center}
        \includegraphics[width=\textwidth]{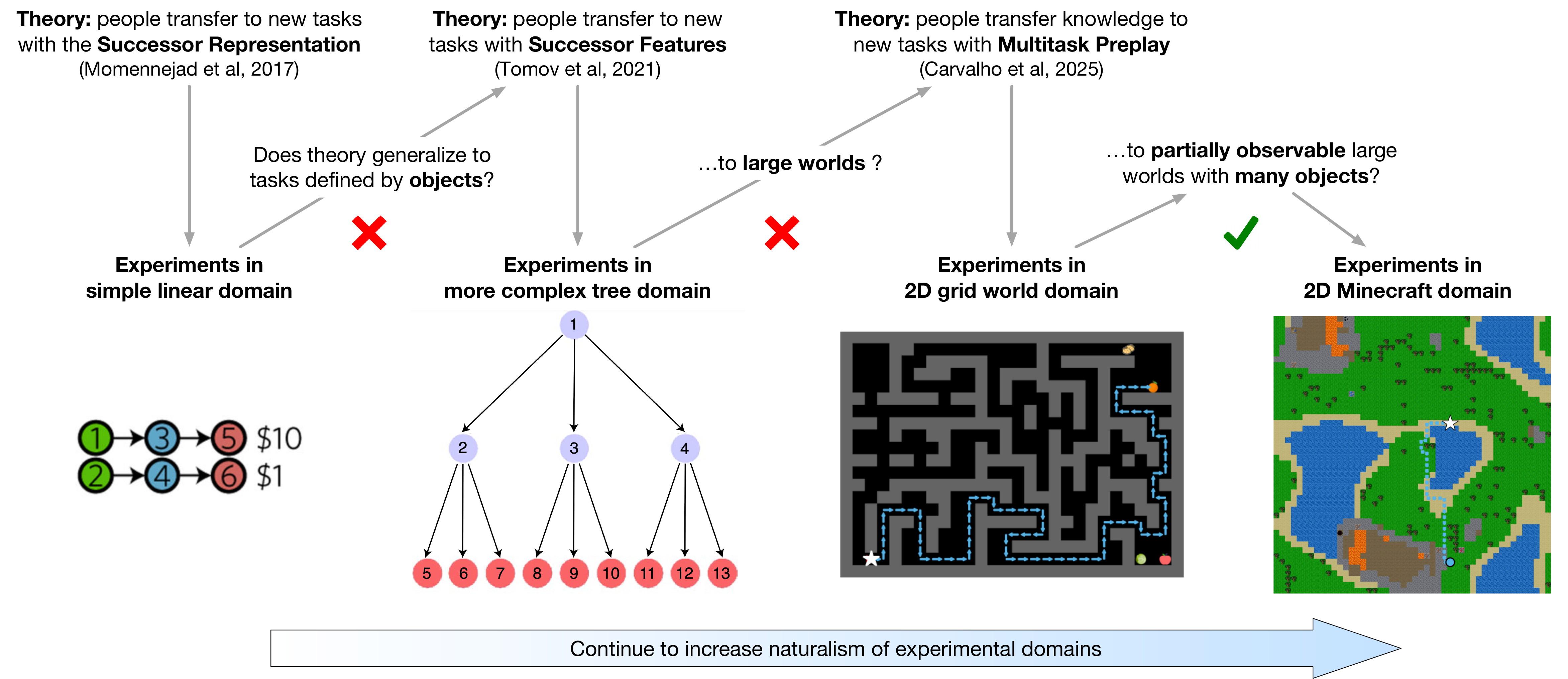}\\
        \includegraphics[width=\textwidth]{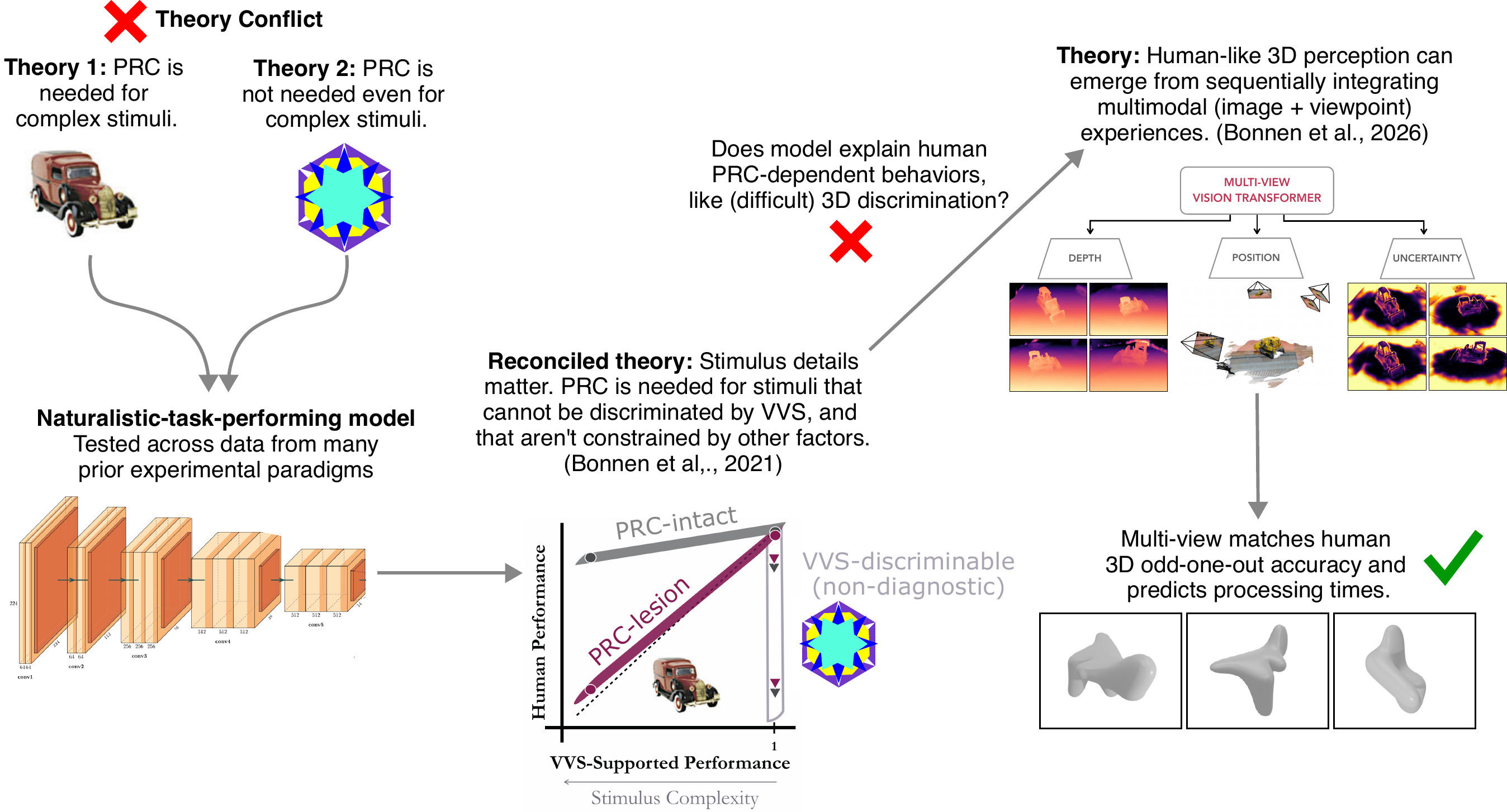}
        \caption{\andrew{Two examples of} \wilka{naturalistic computational cognitive science. Given a computational theory, researchers generate initial predictions to test their theory in a simplified setting. Theories are validated by their ability to predict human behavior in increasingly naturalistic conditions}\andrew{, and refined where they fail to predict correctly. (Top) Progression of theories of sequential task generalization. (Bottom) Examples of progression in understanding PRC involvement in difficult visual judgments.}. Figures adapted from~\citet{momennejad2017successor,tomov2021multi,carvalho2025preemptive,bonnen2021ventral,bonnen2026humanlevel3dshapeperception}.}
        \label{fig:overview-defintion}
    \end{center}
\end{figure}

\wilka{The next two sections address why. We argue that increasing naturalism is necessary---not merely beneficial---for developing a generalizable theory of cognition. In \S\ref{sec:data-benefits}, we review evidence from neuroscience, cognitive science, and AI showing that naturalistic paradigms yield qualitatively different behavior and engage different computational mechanisms---processes spanning visual processing \citep{amme2024saccade}, learning \citep{rosenberg2021mice,collins2024rl}, memory \citep{helbing2020search,hasson2015hierarchical}, and moral reasoning \citep{francis2017simulating} all shift when naturalistic elements are introduced. These findings suggest that naturalistic experiments are essential for building a complete theory of the system. In \S\ref{sec:learning-benefits}, we go further: training on naturalistic data does not merely reveal different mechanisms but can also shape them, producing qualitatively different generalization patterns than training in simplified settings. These differences can alter the inferences we draw from our models and shed new light on the origins of cognitive and neural phenomena. Together, these two sections establish that pursuing naturalistic experiments---and models that accommodate them---is necessary for a complete understanding of cognition (Figure \ref{fig:overview-defintion}).
}

\wilka{The remaining sections address the how. In \S\ref{sec:building}, we distill a key lesson from AI for developing task-performing models in naturalistic settings: building generalizable models requires iterative and collective effort. We propose dynamic meta-benchmarks---living collections that researchers continually expand---as the coordination mechanism, and outline a development strategy for fast, frictionless iteration. In \S\ref{sec:theory}, we show how naturalistic experiments and complex models can anchor generalizable cognitive theories---uniting task-performing predictive models with reductive understanding of underlying mechanisms, using tools such as rational analysis \citep{anderson1990adaptive} and new approaches to interpreting complex models \citep{geiger2021causal}. We close in \S\ref{sec:discussion} with the broader context and implications of naturalistic computational cognitive science.
}

\section{What \textit{is} ``naturalistic'' computational cognitive science?} \label{sec:definitions}

Naturalistic computational cognitive science is a research strategy for theory-driven cognitive science that aims to predict and understand human behavior across increasingly naturalistic settings. This approach gradually increases the ecological validity of experimental designs while ensuring models can accommodate both simplified and more natural conditions. \wilka{Of course, our emphasis on naturalism does not imply that simple experimental paradigms are useless---they remain valuable, and the goal is to extend them with naturalistic settings rather than replace them.} In machine learning (\S\ref{sec:building}), analogous strategies have enabled models that can work with real-world data, albeit imperfectly.

\wilka{Aside from our focus on learning-based methods, this approach is model-class agnostic and compatible with many learning-based frameworks, including Bayesian models~\citep{lee2013bayesian}, RL~\citep{collins2019reinforcement}, predictive representations~\citep{carvalho2024predictive}, and program synthesis~\citep{rule2024symbolic}. The main criterion we advocate for is that models are theory-driven---not black-box foundation models trained to reproduce experimental data~\citep{binz2024centaurfoundationmodelhuman}, though these may be a useful tool for experimental design.}

As in other approaches within computational cognitive science, this approach starts with formulating models and designing controlled experiments to isolate core phenomena. Like prior work, we place an emphasis on \textit{model generalization}, i.e. predicting human behavior on data the model hasn't been exposed to---including data from new experimental conditions or tasks \citep{busemeyer2000model}. Therefore,
\begin{enumerate}
  \item Model parameters are estimated with a ``training'' distribution before being evaluated.
  \item Models should predict human behavior on held-out examples that neither the model nor humans have encountered.
\end{enumerate}

\begin{figure}[!htbp]
   \begin{center}
       \includegraphics[width=.82\textwidth]{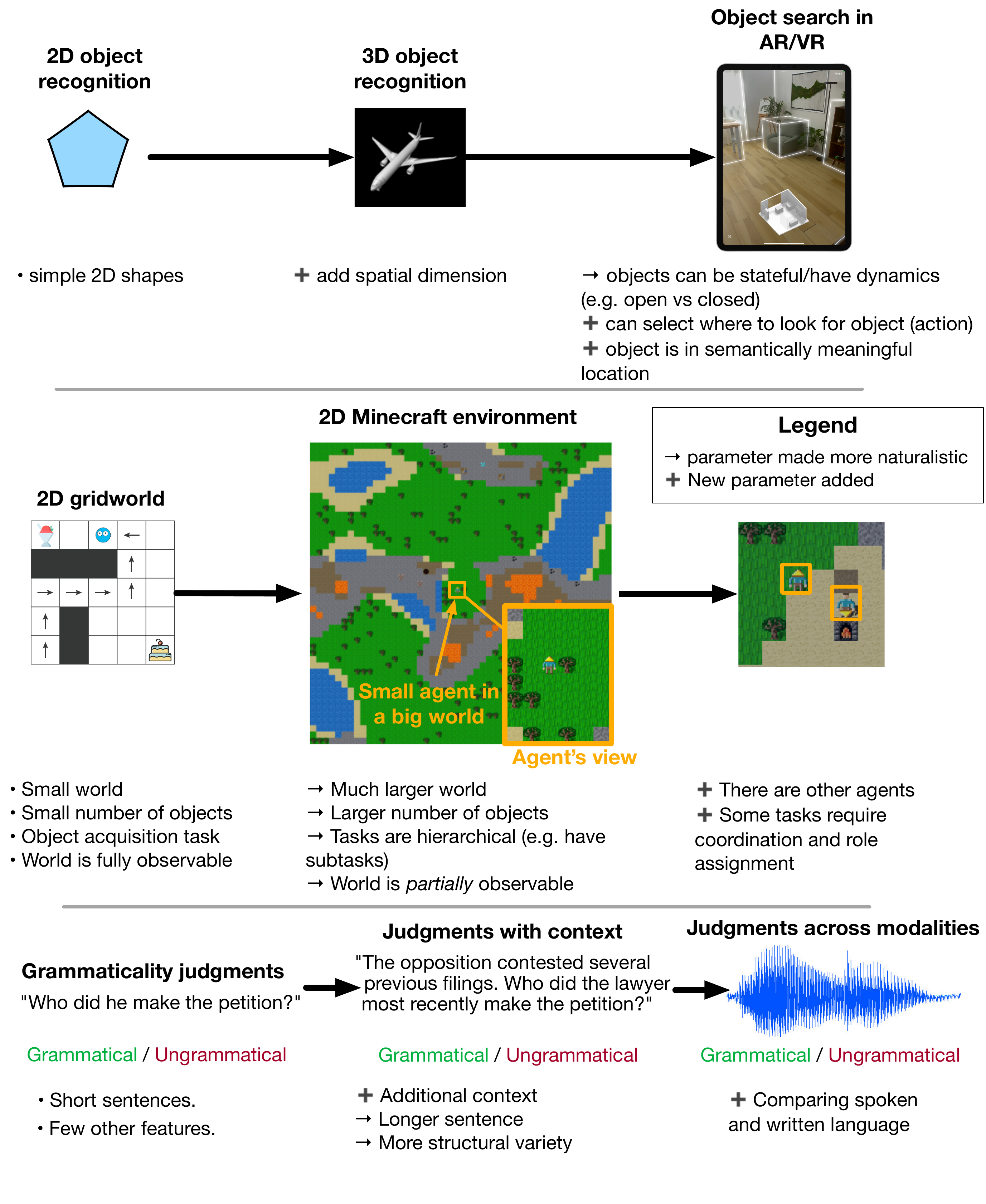}
       \caption{\textbf{Examples of increasingly naturalistic settings that we can now study in a theory-driven manner.} All of these are settings where tasks and stimuli can now be parametrically generated---i.e., thanks to ``generative AI'', we can now \textit{automate} the generation of photorealistic synthetic data and virtual worlds; thanks to virtual and augmented reality, we can now scan and parametrically manipulate real environments. Thanks to progress in AI,  we can now build task-performing models that can operate in these stimuli and task settings. 
       }
       \label{fig:naturalism}
   \end{center}
\end{figure}

One key facet of naturalistic computational cognitive science is that we place emphasis on having models---in particular, the same model---predict human behavior across increasingly naturalistic conditions that cover \wilka{the} space of settings in which the theoretical construct is expected to generalize. We see two main paths toward increasing naturalism in experimental paradigms (and provide examples in Figure \ref{fig:naturalism}):
\begin{enumerate}
    \item Expanding existing experiment parameters to better cover the scope of natural distributions. 
    For example, in an object classification task, we might expand the category set to include the many objects people encounter in daily life.
    \item Adding ecologically motivated parameters that might interact with the theoretical constructs in question. For example, in a linguistic judgment task, we might test judgments across both spoken and written utterances.
\end{enumerate}

We describe a paradigm as more naturalistic if it incorporates a broader scope of the range of natural variability over which the theoretical constructs would be expected to generalize. We describe a model as more naturalistic if it is capable of generalizing its predictions across a broader range of contexts. Note that increasing naturalism does not always mean faithfully approximating all the joint statistics of the natural distribution---increasing the scope of variation along naturalistic dimensions can also enable moving outside the natural data distribution for controlled experiments \andrew{testing a hypothesis} (see \S\ref{sec:theory:controlled_experiments}).

Generally, identifying the parameters for increasing naturalism in a particular experiment is not trivial---it is task- and theory-specific.
To illustrate ``naturalistic'' more explicitly, we therefore offer some examples of dimensions along which naturalism can be increased.

\textbf{Task paradigm}
\begin{itemize}
\item Incorporating multiple paradigms across which a hypothesis would be expected to hold, rather than relying on a single task; for example testing multi-step RL tasks as well as bandit tasks.
\item Incorporating broader stimulus sets, for example augmenting synthetic images with real images, or images with videos.
\item Having task stimuli generated by many varying latent factors, not just the variables of ostensible theoretical interest.
\end{itemize}

\textbf{Environment}
\begin{itemize}
\item Large state spaces that reflect the complexity of real-world environments~\citep{wise2023naturalistic}.
\item Incorporating other social agents that learn with the agent~\citep{hu1998multiagent} and that the organism must interact with~\citep{velez2021learning}.
\item A continually changing environment~\citep{abel2024definition}.
\end{itemize}
\textbf{Model architecture}
\begin{itemize}
\item Architectures that operate over sensory inputs  like natural images \citep{yamins2014performance}, speech \citep{kell2018task}, or natural language \citep{schrimpf2021neural} rather than simplified stimuli (e.g. low-dimensional or discrete inputs containing only task-relevant features).
\item Architectures with naturalistic action spaces---for example, modeling motor control and embodiment rather than abstract, symbolic actions \citep{merel2019hierarchical}.
\end{itemize}
\textbf{Learning algorithm}
\begin{itemize}
\item Unsupervised~\citep{higgins2021unsupervised}, self-supervised~\citep{konkle2022self}, or intrinsic~\citep{chentanez2004intrinsically,oudeyer2013intrinsically} learning algorithms.
\item Social~\citep{henrich2016secret} or cultural~\citep{cook2024artificial} learning algorithms.
\item Prospective algorithms that aim to model how the tasks evolve in continually changing environments~\citep{seligman2013navigating,de2023prospective}.
\end{itemize}

\wilka{These four dimensions are not a definition of naturalism. They are examples of directions along which a theory of cognition must hold up to count as general. Generalizability means extending along axes like these---and often theories are not extended along any.}

\section{The \andrew{importance} of naturalistic experimental settings \andrew{for building generalizable theories}}\label{sec:data-benefits}

\begin{figure}[!htp]
  \begin{center}
      \includegraphics[width=\textwidth]{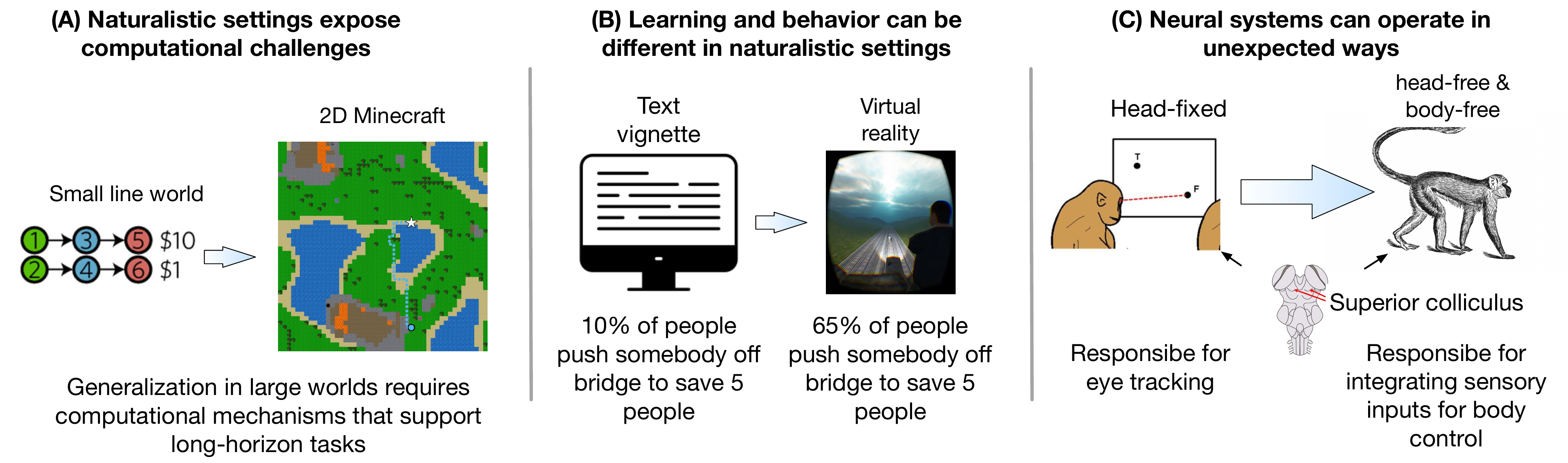} 
      \caption{Overview of the benefits of increasingly naturalistic experimental conditions.}
      \label{fig:overview-experiments}
  \end{center}
\end{figure}

\wilka{As an orienting conceptual example, let us imagine that we are doctors studying heart function, and have chosen to study participants lying at rest. The setting offers high test-retest reliability and a controlled view of processes like how respiration affects heart rate. But that control comes precisely from removing the variation the heart evolved to handle --- rapid adaptation to movement, stress, and the like. Understanding the system in full, and how its physiological and psychological processes interact, requires studying the heart across more naturalistic settings. For instance, manipulating breathing across activities like sitting, speaking, and running would pin down its effect on heart function more sharply than rest alone, and reveal how that effect varies with context. Varied naturalistic settings are how we elaborate the system and build complete theories of its function.}

\subsection{Naturalistic experimental paradigms can expose computational challenges that engage mechanisms differently}

Here, we highlight examples where the challenges posed by increasing naturalism can engage computational mechanisms differently---which can help to disentangle underlying mechanisms, and achieve more generalizable theoretical understanding.

\wilka{\paperarg{Example 1: mechanisms for credit assignment are only exposed when actions have delayed consequences}.
In naturalistic settings, people learn how actions lead to consequences hundreds of timesteps away. Consider navigating a university campus: you learn that heading north leads to the music school, then discover a jazz cafe just off that path. \textit{Credit assignment}~\citep{sutton2018reinforcement} is what lets you recognize the northward route serves both. Prior work modeled this with \textit{successor features}~\citep{gershman2018successor,tomov2021multi}, but only studied human behavior in worlds of around 13 states---small enough that models can succeed without any explicit long-horizon mechanism. In larger worlds, successor features predict fast reaction times but cannot handle nearby-object transfer. In contrast, model-based planning handles this transfer but predicts slow deliberation. Humans display both. To explain this, \citet{carvalho2025preemptive} proposed \textit{multitask preplay}: people mentally rehearse potential detours in advance and cache the results for fast retrieval later on (Figure~\ref{fig:overview-defintion}, top).

Testing multitask preplay required a large environment: in smaller worlds, planning is cheap enough that cached retrieval looks no different from deliberation. Implementing cached retrieval with a brain-plausible learning rule then made the credit-assignment problem concrete---the agent must credit a northward choice (taken to reach the music school) for the cafe discovered along the way. Such cross-task transfer is a known failure mode of temporal-difference learning: approximation errors inflate the value of actions that should not be taken. Multitask preplay required additional machinery to mitigate this, pointing to mechanisms the brain may use for the same job. Because the model was developed in a naturalistic setting, it continued to predict human behavior in an even more naturalistic one---a large, partially observable 2D world with many objects.
}

\paperarg{Example 2: seemingly-similar tasks that engage brain mechanisms differently can be reconciled by naturalistic models}.
\wilka{A long-running debate asks whether perceptual ``oddity'' tasks (identifying the object that differs from the others in a set) recruit the Medial Temporal Lobe (MTL). \citet{murray2007visual} and others argue that MTL is recruited to discriminate ``complex'' stimulus sets that visual cortex cannot resolve alone, so MTL lesions selectively impair performance on those stimuli; others find no such interaction and attribute it to methodological artifacts \citep{buffalo1998human,suzuki2009perception}. \citet{bonnen2021ventral,bonnen2023inconsistencies} resolved the impasse by passing \emph{all} the \emph{raw} stimuli from prior experiments through pretrained convolutional vision models and testing whether the oddity could be picked out in representation space. When the vision model alone could distinguish the oddity, MTL was dispensable; when it could not, MTL was required. Some ``complex'' stimulus sets from prior experiments fell on each side of this split. ``Complexity'' had been underspecified: differences in operationalization that conceptual descriptions abstracted away drove the seemingly-conflicting findings. This example illustrates the value of building models that directly perform the same naturalistic tasks as the subjects, and testing these models on benchmarks that incorporate stimuli from many experimental paradigms---\andrew{resolving otherwise intractable theoretical disagreements.}}

\paperarg{Example \wilka{3}: Different learning mechanisms only yield distinct predictions under working memory load}. 
One challenge of overly-simplified tasks is potential \emph{aliasing} of different solutions, where many different computational approaches yield the same behavior. \citet{collins2024rl} presents one example where a behavior ostensibly implemented as a standard reinforcement learning algorithm may instead be produced from simpler value-free associative learning combined with working memory.\footnote{Whether this alternative counts as an alternative implementation of RL is orthogonal to our argument; our point is that the models yield distinct predictions in only some experimental regimes.} 
Collins notes that ``Disentangling multiple processes requires considering more complex tasks to elicit differentiable behavior.'' The more complex tasks in question simply increase the set of stimuli within learning blocks to a handful of objects, rather than only two or three---i.e., expanding the range of set sizes more clearly disentangles the learning processes. Thus, by restricting to minimal paradigms, we enable a system to use many solutions and prevent ourselves from discriminating between them. By contrast, by imposing a broad range of evaluations on the system \citep[cf.][]{nau2024centering}---especially naturalistic evaluations that increase demands along different axes of task difficulty---the increased constraints actually make it easier to map the model on to natural intelligence. More constraints on the mapping make it less ambiguous, as is highlighted in the \emph{contravariance principle} of \citet{cao2021explanatorypt2}.

\subsection{Behavior can be \andrew{qualitatively} different in naturalistic experiments}\label{sec:benefits:behavior}

\wilka{Cognitive analyses typically isolate a capacity with a task designed to elicit it. But behavior often shifts with seemingly-orthogonal aspects of that task, and theories tuned to one setting risk overfitting to it. Three cases illustrate the risk concretely---putatively well-isolated capacities (memory, learning, moral choice) behave qualitatively differently under naturalistic demands.}

\paperarg{Example 1: memory performance differs between naturalistic search and explicit memorization}.
When participants are explicitly instructed to memorize objects in a 3D home environment, their subsequent recall accuracy was significantly lower than when they incidentally encountered the same objects during visual search tasks~\citep{helbing2020search}.
This effect also extends to spatial memory---participants were able to place objects closer to their original positions in the search condition.
These findings demonstrate that tasks focused explicitly on memory may not capture all aspects of how memory operates during naturalistic behavior.
Thus, computational models of memory may need to be revised to account for these differences in naturalistic settings.

\paperarg{Example 2: mice can learn $1000 \times$ faster during natural behavior, compared to two-alternative forced choice (2AFC) tasks}.
Recent research has shown that learning efficiency can dramatically differ between simplified laboratory tasks and more naturalistic settings~\citep{rosenberg2021mice}.
When mice are tested in traditional 2AFC tasks, they typically require 10,000 training trials over 3-6 weeks to reach asymptotic performance of only about 67\% accuracy.
However, when allowed to \textit{freely explore} a labyrinth (similar to natural burrow systems~\citep{small1901experimental}), mice demonstrate remarkably accelerated learning~\citep{rosenberg2021mice};
they learned to navigate a complex maze with 63 T-junctions in just a few hours, optimizing their path to a reward after only about 10 reward experiences.
\andrew{The time it took them to learn} 
in the naturalistic setting was approximately $1000\times$ faster than in 2AFC paradigms.
\andrew{While it may not be meaningful to directly compare ``learning rate'' as a parameter between the settings, this clear and }
dramatic difference in learning \andrew{efficiency} demonstrates how traditional simplified experimental paradigms may fundamentally underestimate an organism's learning capabilities, and suggests that learning mechanisms may operate differently when they are engaged through more naturalistic behavior.

\paperarg{Example 3: moral behavior can change when moving from textual vignettes to virtual reality}.
Recent research shows that behavioral choices can change when moving from text to virtual reality, by driving a shift from moral \textit{judgment} (i.e. deciding whether things are moral) to moral \textit{action}~\citep[][see Figure~\ref{fig:overview-experiments} A]{francis2017simulating}.
When participants were presented the classic trolley dilemma in text form, only 40\% said they would pull a switch to divert the trolley, thus killing one person to save five others.
However, when the same scenario was presented in virtual reality, where participants must simulate the action, 55\% chose to pull the switch.
The difference was even more pronounced in the ``footbridge'' variant of the dilemma.
In the text version, only 10\% of participants said they would push someone off a bridge to save five others.
However, in virtual reality, 65\% of participants choose to push the person.
These findings demonstrate that people's stated moral preferences in hypothetical scenarios can differ substantially from their actions in more naturalistic settings.

\subsection{Neural systems can operate differently under naturalistic conditions}

In addition to behavioral differences like those reviewed above, naturalistic conditions can alter neural coding and computation.
In this section, we highlight some examples where different computations or roles arise when moving from standard paradigms to more naturalistic ones.

\paperarg{Example 1: the role of superior colliculus in eye movements vs. integrating sensory inputs for body control}.
\citet{cisek2024toward} argues that as we increased naturalism from head-fixed to head-free and body-free settings for studying monkeys, we expanded our theory of superior colliculus from controlling saccadic eye movement to generally integrating multimodal cues to guide bodily action-selection (Figure~\ref{fig:overview-experiments} B). That is, as we increased the naturalism of the experimental conditions, we arrived at a more complete model of superior colliculus.
This illustrates the more general point, that overly focusing on computation as an input-output mapping neglects the fundamental fact that natural intelligence evolved not for single responses, but for closed-loop control in an environment \citep{cisek1999beyond}. This perspectives yields a different interpretation of the system's representations and processes. More generally, many researchers argue that cognition cannot be understood completely outside its embodiment and environment \citep[e.g.,][]{newen2018oxford}.

\paperarg{Example 2: visual processing differs during natural viewing compared to passive viewing paradigms}.
One early visual response in the brain (the P100/M100) occurs about 100ms after a person's eyes fixate on a new location~\citep{vinje2000sparse}.
However, this finding comes primarily from experiments where participants passively view images while fixating.
In contrast, natural vision involves frequent active eye movements to sample information throughout a scene.
When \citet{amme2024saccade} examined brain responses during more natural active viewing, they found that the P100/M100 response actually begins when the eye movement (saccade) \emph{begins}, not when it ends at a new fixation point.
Furthermore, the neural patterns during active viewing were \emph{anti-correlated} to those seen during passive viewing, suggesting fundamentally different processing mechanisms.
These findings reveal that simplified experimental paradigms, while valuable, may not capture all aspects of how visual processing operates during natural behavior.

\paperarg{Example 3: model-based systems are used for learning moral judgments, but model-free systems are used for learning to avoid harming others}.
Despite moral reasoning being strongly associated with model-based learning in the prefrontal cortex, neural evidence shows a surprising shift when people actually learn to avoid harming others~\citep{lockwood2020model}. Moral reasoning commonly engages model-based systems in the lateral prefrontal cortex (LPFC)~\citep{spitzer2007neural,carlson2018lateral} with LPFC disruption leading to reduced norm compliance and enforcement~\citep{knoch2006diminishing,ruff2013changing}. However, when participants perform tasks requiring them to learn moral behavior (rather than just reason about it), researchers instead observed neural activity consistent with model-free learning. As in \S\ref{sec:benefits:behavior}, this illustrates that when we increase the naturalism of experiments (learning to take moral actions rather than simply reasoning about them), neural systems can operate in unexpected ways.

\section{The (surprising) benefits of learning with naturalistic data}\label{sec:learning-benefits}
\begin{figure}[htp]
    \begin{center}
        \includegraphics[width=\textwidth]{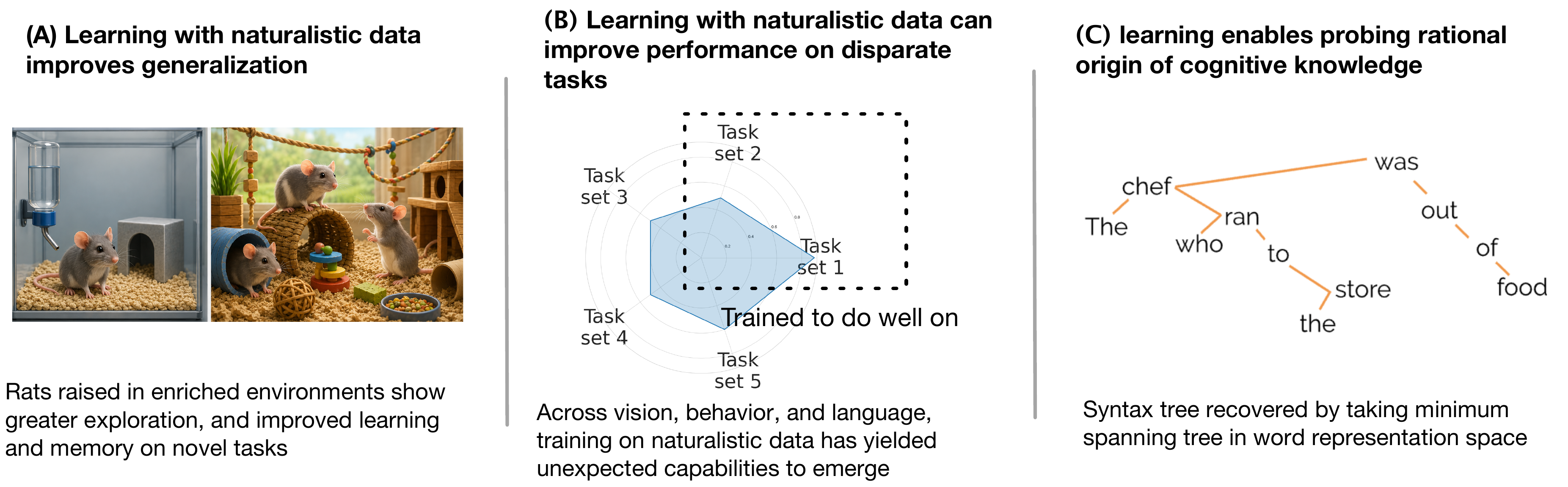} 
        \caption{Overview of the benefits of learning with naturalistic data. (Figures reproduced from \citealp{manning2020emergent}.)}
        \label{fig:overview-learning}
    \end{center}
\end{figure}
A standard lesson in science is that we must simplify an experimental setting to better arrive at a causal conclusion \citep[e.g.][]{rust2005praise}. 
However, we have reviewed examples illustrating that more is different \citep{anderson1972more}---some properties of intelligence may only emerge in more ``complicated'' naturalistic settings.
Here, we detail examples where naturalistic data \emph{itself} plays an important role in producing the empirical phenomena of intelligence.

Many results we discuss are drawn from modern AI research. 
One common trend is a positive interaction between learning-based systems and naturalistic data---learning-based systems can accommodate naturalistic data, and reciprocally, learning from naturalistic data results in qualitatively different generalization than learning in simpler settings. 
While we do not yet know the full implications for natural intelligence, these results open interesting questions and challenge prior assumptions.
In particular, cognitive science should consider the limitations placed on our models\andrew{---and correspondingly, on the theories we instantiate in those models---}when we fail to consider naturalistic data.

\subsection{Learning from naturalistic data can improve generalization}

Studying how systems generalize is fundamental to \andrew{building theories in} cognitive science and AI. 
\andrew{In order to study generalization in simplified settings, we need to make the key assumption} 
that the simplifications do not alter the generalization problem. Here, we review studies from AI and neuroscience showing that variability in experience (even along seemingly-orthogonal axes) can alter learning and generalization---thus implying that studying models within more restricted settings may mislead us about the naturalistic computational problem. 

\paperarg{Example 1: Compositional generalization of image classifiers vs. agents}. Naturalistic tasks can fundamentally change what models learn and how they generalize. \citet{hill2019environmental} trained two models on a vision-language grounding task, and tested their compositional generalization to held-out instructions. Both models were trained on the same language set, and tested on the same held-out examples. However, one model was trained as an agent interacting with a simulated environment, while the other was simply trained to classify screenshots from that environment based on language instructions. Surprisingly, the interacting agent exhibited perfect compositional generalization to novel language utterances, while the image classifier generalized substantially worse. The authors also found generalization benefits of other naturalistic factors, including 3D (rather than 2D) environments, or egocentric embodiments. These results illustrate how richer naturalistic settings can enhance the generalizability of learned solutions. Thus, if we are interested in understanding generalization, we may need to build models that learn from appropriately naturalistic data.

\paperarg{Example 2: Syntactic generalization is enhanced by variability in other structures' fillers}
Structural generalization is of deep interest in linguistics. Recently, \citet{misra2024language} performed controlled experiments in which challenging linguistic constructions are systematically held out from language model training data, and showed that models trained on naturalistic language generalize to held out constructions by composing pieces of simpler constructions. Critically, this generalization depended on sufficient variability in the semantic fillers observed in the structures seen in training---thus, incorporating naturalistic variation in model training data can impact our theoretical inferences about generalization.

\paperarg{Example 3: Rats raised in enriched environments}.
Analogously, neuroscientists have found that animals raised in more complex environments can be more skilled than those raised in simpler environments. 
Rats raised in enriched environments --- with social interaction and/or more space or toys --- show greater exploration, and improved learning and memory on novel tasks \citep{simpson2011impact}. Thus, for natural as well as artificial intelligence, greater naturalistic variation in experience can alter the system's learning in ways that impact our experimental and theoretical inferences.

\subsection{Learning with naturalistic data can yield good performance across a range of seemingly disparate tasks}
One interesting finding from AI is that when highly-parameterized deep learning architectures are trained with bountiful naturalistic data and appropriate ``basic'' learning objectives, these architectures can develop mechanisms that go beyond the training objective. Crucially, these learning paradigms can allow models to transfer well (through initial performance or accelerated learning) on novel tasks beyond the training distribution.

Human learning may likewise benefit from transfer among the disparate tasks we learn. 
We argue that these AI findings should motivate cognitive science research studying whether ``downstream human behavior'' on a set of tasks can be recapitulated by a model which is trained on varied naturalistic tasks or stimuli representative of (some aspects of) human experience.
Below, we provide examples illustrating this kind of transfer.

\paperarg{Example 1: Computer Vision}.
One early success of deep learning came in computer vision. Soon after AlexNet~\citep{krizhevsky2012imagenet} achieved strong results on the ImageNet dataset~\citep{deng2009imagenet}, researchers showed that AlexNet's features could be repurposed to novel tasks like scene recognition or medical diagnosis~\citep{donahue2014decaf,sharif2014cnn,litjens2017survey}. This result was striking because features trained via supervision on fixed object classes nevertheless enabled transfer to novel objects or even different types of tasks. This pattern has continued, and more recently researchers showed that the representations learned by these models could transfer from ImageNet to 3D household navigation and dexterous tasks, sometimes achieving \textit{better} performance than using ``ground-truth'' state~\citep{parisi2022unsurprising,yuan2022pre}.
Beyond their useful representations, vision models trained with naturalistic data also develop mechanisms they were not \andrew{explicitly} trained for.
For example, scene-oriented CNNs develop mechanisms for object detection~\citep{zhou2014object} and vision transformers develop mechanisms for segmentation~\citep{caron2021emerging}---both without an explicit training signal.

\paperarg{Example 2: Reinforcement learning}. Likewise in reinforcement learning, researchers have found that agents trained on many (billions) of tasks exhibit strong ``out-of-distribution'' generalization on unknown tasks~\citep{team2021open}, learning novel tasks as efficiently as humans~\citep{team2023human}. Some algorithms can even generalize to collaborating with humans without any human data~\citep{strouse2021collaborating}. Recently, reinforcement learning algorithms for learning a modern variant of the successor representation~\citep{gershman2018successor,carvalho2024predictive} have been shown to develop skills for exploration and behavior without explicit training signals~\citep{liu2024single}.

\paperarg{Example 3: Large language models}. Perhaps the most striking example of this general phenomenon comes from large language models. These models are only trained to predict the next ``token'' (or word) to appear in a sequence. However, the distribution of internet language effectively includes a broad mixture of many tasks \citep{radford2019language}. When trained on these vast naturalistic data, language models develop mechanisms for disparate tasks such as modular arithmetic, solving word problems, etc.~\citep{wei2022emergent}---and even for adapting to new tasks from examples \citep{brown2020language}.
Moreover, the representations learned by these models can transfer to superficially dissimilar downstream tasks \citep{lu2022frozen}, including playing video games \citep{reid2022can}. These examples illustrate how learning from a broad naturalistic distribution can induce many abilities---and can transfer to downstream tasks. 

\subsection{Learning from naturalistic data allows us to ask new questions about the origins of knowledge}

The points outlined above have important consequences: using models that learn from naturalistic data can change our theoretical conclusions in cognitive science. For example, if a model fails to generalize from simple data, we cannot be sure whether the model or the data are inadequate. By contrast, using models that learn from naturalistic data can enable us to ask more precise questions about which features---of models or data---are necessary to reproduce the theoretically-relevant features of cognitive and neural processes.

\paperarg{Example 1: language models and learnability}.
Recent language modeling \andrew{works raise} challenges to prior assumptions about language innateness.\footnote{We briefly sketch these issues; see \citet{piantadosi2023modern} and \citet{futrell2025linguistics} for thorough accounts.} Classical approaches to language focused on simple models \citep[e.g.][]{chomsky2014minimalist} \andrew{under strong assumptions. Using certain assumptions, \citet{gold1967language} proved that even relatively simple languages cannot be learned from input alone. This contributed to arguments that language's precursors must be innate \citep[e.g.]{chomsky1957syntactic,chomsky1965aspects}---which influenced cognitive science writ large \citep{fodor1975language,fodor1988connectionism,quilty2023best}. }
However, these theories did not attempt to model language processing in all its naturalistic detail.

\andrew{By contrast, deep-learning models that learn from plentiful naturalistic data acquire} aspects of syntax \citep{manning2020emergent} and semantics \citep{li2021implicit}---and capture behavioral and neuroscientific phenomena \citep{schrimpf2021neural}. \andrew{Although these models often learn from inhuman quantities of language \citep{wilcox2024bigger}, even human-like quantities allow learning complex syntactic features that were previously considered evidence for innate structure \citep[e.g.][]{wilcox2023using}.
Furthermore, controlled experiments---like \citet{misra2024language}, reviewed above---show how naturalistic variation contributes to this syntactic generalization.} Thus, studying learnability from naturalistic data has helped to reshape our understanding of the necessary and sufficient features for acquiring linguistic capabilities.



\wilka{\paperarg{Example 2: naturalistic data can lead to brain-like functional specialization}.
Deep neural networks trained on natural images have long predicted neural responses from V1 through IT~\citep{olshausen1996emergence, lee2007sparse, yamins2016using,doshi2023cortical}, establishing them as candidate models of visual processing. Recently, researchers have found that training with naturalistic data can yield brain-like functional specialization---something once thought to require innate hard-wiring.} 
Given the social importance of faces, it might seem likely that face recognition is innately encoded in the brain. Indeed, the discovery of the Fusiform Face Area (FFA)---a visual region specialized for face perception \citep{kanwisher2006fusiform}---supports this view. However, \citet{dobs2022brain} showed that this specialization can emerge from a domain-general vision model trained on a naturalistic distribution of object and face recognition tasks.
As the authors state: ``It may be that the only inductive bias humans need to develop their face system is the already well-established early preference of infants to look at faces.''
Of course, these results do \emph{not} necessarily imply that the brain's functional organization is not innately specified. However, they illustrate how training computational models on naturalistic data can generate new hypotheses about the origins of neural organization.

\paperarg{\andrew{Example 3: naturalistic multi-view signals can yield human-level 3D shape perception}}. 
\andrew{While large vision models trained on classification tasks capture some features of human-like vision, they show subhuman performance in areas like 3D object perception (\citealp{bonnen2024evaluating}; cf. \citealp{malhotra2023human}). However, recently \citet{bonnen2026humanlevel3dshapeperception} suggest that this may be due to these models lacking another key feature of human visual experience: the experience of seeing a scene from multiple angles (due to movement or stereo vision). Indeed, the authors show that models that are trained to predict view positions from multiple views of the same scene can match human-level 3D shape perception accuracy---thus showing how naturalistic learning signals beyond classification or image-only self-supervision might contribute to human 3D perceptual abilities.}

\section{Building generalizable models \wilka{for naturalistic settings\newline through systematic and collective effort}} \label{sec:building}

\begin{figure}
  \begin{center}
      \includegraphics[width=\textwidth]{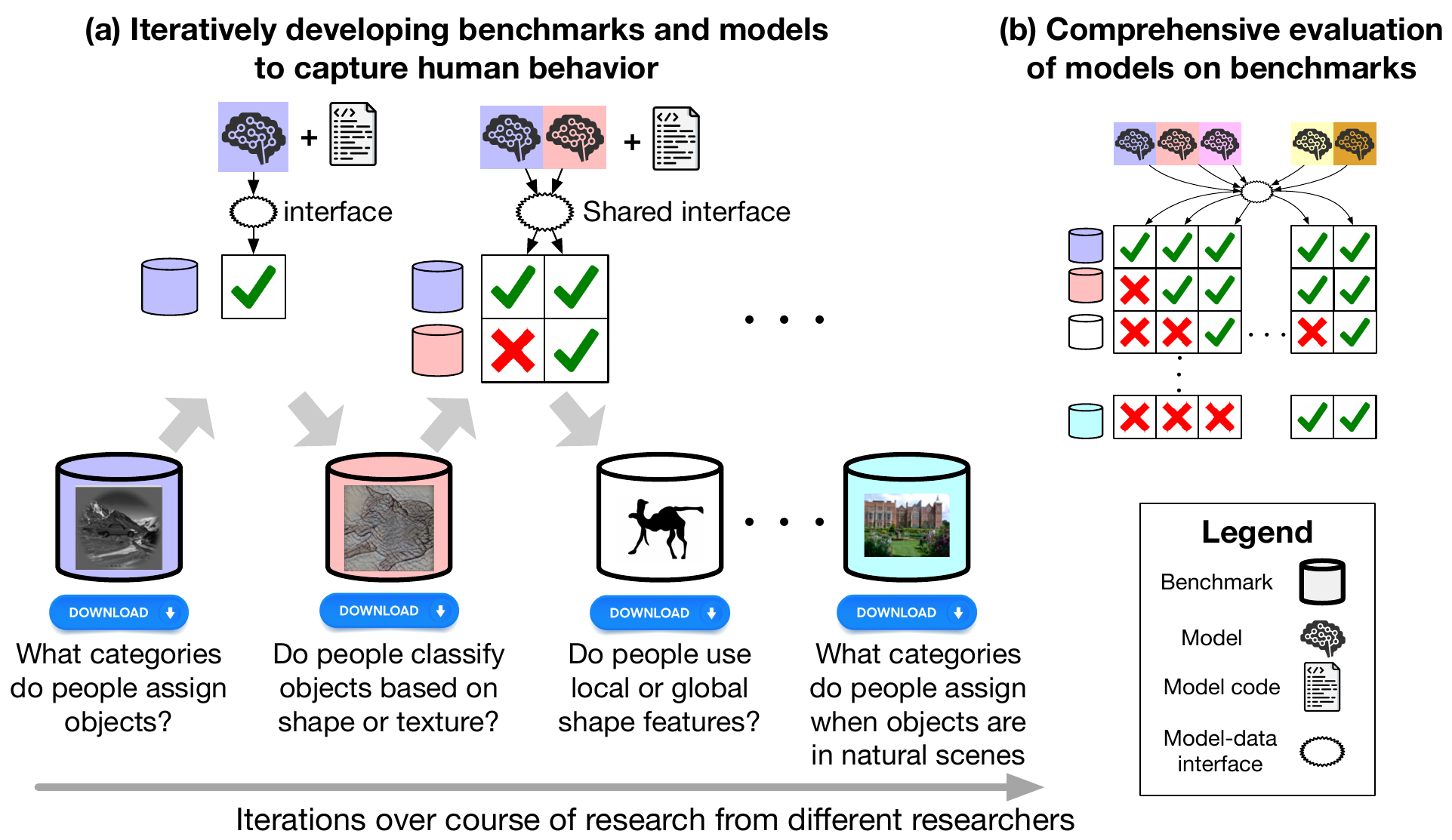} 
      \caption{
\textbf{How cognitive scientists can systematically build generalizable models of cognition that explain human behavior on naturalistic data}.
Our key proposal is leveraging \textit{dynamic meta-benchmarks} (\S\ref{sec:build-benchmarks}) where (a) individual benchmarks aggregate human data from experiments studying facets of a cognitive capacity. The benchmarks evolve as researchers add new facets---for example, extending object classification benchmarks to test whether classification relies on shape versus texture, then making stimuli (at least partially) reflect real-world data distributions. (b) Models are developed with \textit{generalization} as the primary goal (\S\ref{sec:build-generalization}): the same model should predict human behavior across all benchmark facets. (c) To facilitate \textit{frictionless reproducibility} (\S\ref{sec:build-frictionless}), models are evaluated through a standardized interface. (d) This makes it easy for researchers to comprehensively evaluate their hypothesis (i.e. model) against a large swath of human behavioral data capturing different facets of a cognitive capacity. The best hypothesis is then the one that captures human behavior over the most human data. \citep[Images reproduced from][]{geirhos2018imagenet,bowers2023deep}
}\label{fig:overview-engineering}
  \end{center}
\end{figure}

\wilka{


How do we build generalizable models that predict and explain human behavior in naturalistic settings? ML has made substantial progress in building models that exhibit aspects of intelligence in settings closer and closer to the real world. We argue that this trajectory holds lessons for cognitive scientists who aim to model human-like intelligence in such settings.

The first lesson is that progress requires \textbf{systematic} iteration from simpler settings to more naturalistic ones. In principle, a single researcher can systematically cover the space of experiments needed to build generalizable models. In practice, the design space is vast. This makes individual progress difficult. The second lesson is that \textbf{collective effort} enables fast progress over a vast design space. Initially, ML researchers worked on disparate, self-defined problems, leading to diffuse progress~\citep{recht2024mechanics}. Progress accelerated when the field adopted \andrew{shared} benchmarks. 
A widely adopted benchmark is a sociological object: it certifies, often through peer review, that experts judge a problem worth solving.
With the right infrastructure, this allows large populations of researchers to make fast progress together.

\textbf{We propose that cognitive science adopt dynamic meta-benchmarks}: living collections of benchmarks that researchers \andrew{continually} contribute to, where each benchmark tests a specific facet of a cognitive capacity (\S\ref{sec:build-benchmarks}, see Figure \ref{fig:overview-engineering}). Consider object recognition. One facet might test whether classification relies on shape or texture; another whether shape processing uses local or global features; a third how classification interacts with scene context. Meta-benchmarks grow in two ways: researchers can increase the naturalism of existing facets \citep[e.g., expanding shape-versus-texture stimuli from simple shapes to shapes derived from naturalistic stimuli;][]{geirhos2020shortcut}, or add new facets exposing untested dimensions such as whether objects are classified by local- or global-shape features~\citep{baker2020local}. \andrew{In either way, benchmarks can cumulatively incorporate experimental tests of new hypotheses, without relinquishing the demand to explain prior results.} A model that predicts human behavior across the full meta-benchmark provides evidence for a generalizable account \andrew{of human capacities within that domain}; failures on particular benchmarks identify where \andrew{an} account breaks down. 

Adopting meta-benchmarks raises an engineering problem: how \andrew{can} we develop models that can operate across many benchmarks, stimulus sets, and task sets? The key is to \textbf{develop models with an eye towards generalization} by ensuring both that models generalize to new data, and that their \textit{development procedure} generalizes as well (\S\ref{sec:build-generalization}). That is, the procedure used to develop a model for one benchmark should allow development for other benchmarks with minimal modification. This is essential for efficiently validating models across a full meta-benchmark.

But how do we adopt development strategies that promote reuse across many benchmarks, stimulus sets, and task sets? One key strategy is \textbf{frictionless reproducibility} ~\citep[\S\ref{sec:build-frictionless};][]{donoho2024data}, where researchers develop research artifacts that can be reused and repurposed with minimal effort.
This includes standardized evaluation protocols and common interfaces that allow researchers to easily run any model against any benchmark. 
We argue that together, these pieces will enable cognitive scientists to develop generalizable models that capture the full range of natural behavior for a cognitive capacity in question.
}

\subsection{Benchmarks: catalysts for progress and innovation}\label{sec:build-benchmarks}

\wilka{Since at least \citet{newell1994unified}, cognitive science has sought to develop unified, task-performing models that explain cognition as a whole~\citep{kriegeskorte2018cognitive,almaatouq2024beyond}.
While cognitive scientists are aware that relying on a single paradigm may reduce the generalizability of conclusions~\citep{holzmeister2024heterogeneity,yarkoni2022generalizability}, the field currently lacks infrastructure for this kind of multi-paradigm evaluation. 
The result is fragmentation and Newell's critique---that piece-wise investigation of brain components would not yield a cohesive understanding~\citep{newell2012you}---remains relevant forty years later~\citep{schrimpf2020integrative,dicarlo2023let}; findings from different experimental paradigms are difficult to build upon and resist integration, even within a domain~\citep{almaatouq2024beyond}.
We discuss how two concepts from ML---benchmarks and leaderboards---can enable collective, systematic effort within cognitive science.}

%
%

\subsubsection*{Recommendations for cognitive science}

\paperarg{R1. Embrace the leaderboard principle}.
A common concern with benchmarks is that researchers will ``overfit'' to test sets, which theoretically should underestimate the true test error \citep{hastie2009elements}. Yet despite widespread iteration on benchmarks in machine learning 
the field has produced models with strong performance on held-out data. This apparent paradox is partially explained by a phenomenon known as the ``leaderboard principle''~\citep{hardt2022patterns}. Researchers typically only publish and build upon models that demonstrate substantial improvements over prior results, rather than minor variations. This selective pressure towards meaningful advances effectively constrains the degree of adaptation to test sets, as researchers focus on substantial improvements rather than exhaustively exploring the test set's peculiarities. 
Nevertheless, a risk of overfitting to data remains. We argue that cognitive science can avoid this through the creation of dynamic phenomena-oriented benchmarks, where researchers can contribute new challenges that identify a missing component of a \andrew{cognitive domain} of interest. We detail this below.

\paperarg{R2. Develop \textit{dynamic} benchmarks to
iteratively capture important phenomena}. 
ML benchmarks are not collected to test specific hypotheses~\citep{hardt2022patterns} and so may seem unscientific.
However, cognitive science can embrace scientific benchmarks \andrew{that allow} testing specific hypotheses, \andrew{while situating those tests within the context of the broader phenomena of the cognitive domain}. 
We use the story of ``Brain-Score'' as an illustrative example of \textit{initial} steps in this direction.

Brain-Score~\citep{schrimpf2020integrative} began as a benchmark collecting behavioral and neural data from humans and monkeys on object recognition tasks.
One common critique of Brain-Score is that \andrew{it does not test \textit{hypotheses} about human object recognition.} 
For example, \citet{bowers2023deep} noted that matching human predictions is insufficient to identify a good model of human object recognition.
The authors reference 9 experiments from the psychophysics literature which presented stimuli generated to test specific hypotheses\andrew{---for example, testing use of local shape features~\citep{baker2020local}, shape vs. texture ~\citep{geirhos2018imagenet}, and contours ~\citep{puebla2022can}---}and showed that deep neural networks failed to capture human behavior across these.
In response,~\citet{dicarlo2023let} added many of these stimulus sets to Brain-Score, so that future models can be evaluated against all of them.
This story illustrates what an exciting strand of computational cognitive science research could look like: one where researchers develop models and experiments that easily integrate with prior findings on a phenomenon thanks to phenomena-oriented benchmarks, generalizable models (\S\ref{sec:build-generalization}), and frictionless reproducibility (\S\ref{sec:build-frictionless}).

We are not the first to identify that cognitive science is fragmented and to suggest benchmarks that aggregate results from the literature as a solution. Indeed, researchers spanning memory, timing, confidence and more have already begun to do so~\citep{macwhinney1985child,oberauer2018benchmarks,aydougan2024timing,rahnev2020confidence,frank2017wordbank}.
Here, we both encourage more subdisciplines within cognitive science to follow suit\andrew{, and} emphasize the importance of frictionless reproducibility and building generalizable models. While prior work has begun \andrew{developing} benchmarks, documentation for downloading and using the data is sometimes lacking; download access is not always authentication-free; and maybe most importantly, if one wants to add new data, there is not always an easy way to re-run all previous models and ensure that findings are commensurate. 
\andrew{More c}ritically, while prior benchmarks have focused on aggregating experimental results from ``simplified'' experiments, we advocate for dynamic benchmarks whose development is oriented around explaining data from increasingly naturalistic experiments (i.e. experiments with either parametrically generated naturalistic stimuli or, eventually, stimuli taken from real-world conditions).


\subsection{Developing models with an eye towards generalizability}\label{sec:build-generalization}

\wilka{Ideally, the models we develop generalize beyond the setting they were designed for.
This requires two things.
First, that the model is \textit{generalizable}, i.e. can learn from, operate over, and perform new tasks.
Second, that the behavioral (and potentially neural) predictions made by our models capture data on these new tasks.
Clearly, the first \andrew{requirement} is a \andrew{prerequisite} for the second.
Here, we detail suggestions from ML for how cognitive science can develop more generalizable models.
}

\subsubsection*{Recommendations for cognitive science}

\paperarg{R3. Iterate between (1) model-design (2) task design where the other is fixed}.
One crucial lesson from ML research is the importance of separating model development from task design.
\citet{patterson2023empirical} strongly caution against simultaneously developing both the problem setting (datasets or tasks) and the solution method, as researchers may inadvertently design evaluations that favor their proposed solution.
Historical examples illustrate this risk. When developing Q-learning algorithms, researchers modified the classic pendulum control problem by adding episode cutoffs and random state resets \citep{machado2018revisiting}. These modifications made Q-learning more tractable by implicitly improving exploration, but thereby obscured fundamental limitations in Q-learning's exploration compared to policy gradient methods.
Maintaining a strict separation between task design and model development enforces a clear delineation between hypothesis formation and testing, which is \textbf{important for reducing experimenter bias}.
This is especially important because researchers tend to find ``positive'' results~\citep{scheel2021excess,sarafoglou2022survey}, a bias exacerbated by an increasingly competitive academic landscape~\citep{lee2014publish, reithmeier201910, woolston2021researchers}.
Much of ML's progress can be attributed to researchers evaluating their models on independently developed datasets and tasks---a practice that both simplifies the research process and promotes scientific rigor.

Thus, we recommend that cognitive science also develop models on tasks/stimuli developed by others.
This may seem to contradict \andrew{the scientific method}, where researchers craft experiments to test their specific hypotheses. The data from prior work was for a different hypothesis!
\andrew{However, this is precisely where dynamic, phenomena-oriented benchmarks can offer a solution. First, where researchers hypothesize a failure of existing theories or models, they can develop a new experimental paradigm that illustrates it, and then commit this paradigm to a dynamic benchmark. Once the paradigm has been publicly committed, they can move on to proposing and testing new models to explain it---which will simultaneously be tested for their continued ability to explain the other phenomena included in the benchmark. In this way, researchers can pursue hypothesis-driven experiment- and model-development, while maintaining scientific rigor by avoiding the temptation to overfit tasks and models to each other.}
We detail \andrew{the practical approaches that enable this beneficial cycle} further in the next section.

\paperarg{R4. Develop models with simplified tasks, but ensure they work with increasingly naturalistic tasks}. The frame problem~\citep{shanahan2004frame} and the ``simulation is doomed to succeed'' critique~\citep{grim2013simulations} flag a recurring failure: the evaluation tasks we design rarely capture the full complexity of the phenomenon we care about. ML's response is to demand that models succeed in both simplified and increasingly complex settings. RL exemplifies the trajectory: initial environments were grid-worlds~\citep{sutton2018reinforcement}, later 2D video games~\citep{bellemare2013arcade}, then virtual 3D homes~\citep{szot2021habitat,li2021igibson}, and now open-world environments~\citep{matthews2024craftax,hughes2024open}. Even with this push toward naturalism, researchers still rely on simpler settings to study specific aspects of model behavior and generalization~\citep{osband2019behaviour,obando2024small}. Indeed, the standard strategy is to \textit{first} develop a model in simplified settings, then verify that it continues to work in more naturalistic ones. This has yielded both theoretical~\citep{grimm2020value,lyle2022understanding} and empirical~\citep{obando2024small,farebrother2024stop} understanding of how RL models work, alongside performant models for complex settings---beating world experts in Go~\citep{silver2017mastering} and controlling real robots~\citep{levine2016end,cheng2024extreme}.

\paperarg{R5. Develop models that work on many tasks}.
The ultimate goal of ML is generalization~\citep{highleyman1959generalized,hardt2022patterns,recht2024mechanics}.
Initially, this meant models that generalize to new data.
Now, it means both models and learning procedures that generalize to new settings.
Thus, evaluating models across diverse datasets has become foundational in ML.
Early examples include ``dropout''~\citep{srivastava2014dropout}, which demonstrated improvements across 6 image and speech datasets, and MoCo~\citep{he2020momentum}, which validated its approach on 9 vision datasets. 
Another notable example is the seminal ``Deep Q-Network'' paper~\citep{mnih2015human}, which showed that a single neural network architecture with fixed hyperparameters\footnote{E.g., the number of layers or the learning rate.} could master 50 different Atari games.
The field has since evolved to organize environments by the cognitive capabilities they test (e.g., exploration, generalization, manipulation)~\citep{patterson2023empirical}, with researchers developing models on some task sets and validating on entirely different ones to ensure robust generalization~\citep{machado2018revisiting}.

We recommend cognitive scientists also test the models that support their theories more broadly. Importantly, we should evaluate models on different tasks from those we use to develop them~\citep{machado2018revisiting}. This further helps prevent our model (and thereby theory) from ``overfitting'' to a particular paradigm.


\subsection{Frictionless reproducibility: a superpower}\label{sec:build-frictionless}
\wilka{Cognitive science and psychology face substantial friction reusing prior data and code~\citep{nosek2015promoting,hardwicke2018data}: data are rarely standardized, analyses are often underreported, and shared code frequently contains errors~\citep{poldrack2017scanning}. In a large-scale replication analysis, $38\%$ of code was not usable and only $31\%$ of workflows reproduced~\citep{hardwicke2018data}. Meanwhile, data-centric sciences are undergoing a profound transition---what \citet{donoho2024data} calls \textit{frictionless reproducibility}---in which good data- and code-sharing practices combine with competitive challenges to let researchers re-execute prior workflows with minimal effort, accelerating progress dramatically. ML, Donoho argues, is ostensibly the \textit{most successful} at frictionless reproducibility. Below we review some strategies that cognitive science can adopt. \S\ref{sec:build-benchmarks} addresses competitive challenges via dynamic meta-benchmarks; here we focus on data and code sharing.
}

\subsubsection*{Recommendations for cognitive science}

\paperarg{R6. Use standardized evaluation protocols where models are compared via a common interface}.
Machine learning's foundation of standardized evaluation protocols traces back to 1959, when~\citet{highleyman1959generalized} created the first alphanumeric pattern-recognition dataset and the first ever ``train-test'' split.
This strategy continued with the famous MNIST dataset~\citep{lecun1998gradient}. MNIST simplified data-sharing and evaluation by pre-processing data, fixing train and test distributions, and sharing the data online.
This standardization enabled fair comparisons between diverse approaches, from boosted stumps~\citep{kegl2009boosting} to support vector machines~\citep{cristianini2002support} or nearest neighbors~\citep{khan2017mcs}.

Standardized evaluation protocols serve a critical function: \textbf{mitigating personal bias} in model assessment. The evolution to Caltech 101~\citep{fei2004learning} and ImageNet~\citep{deng2009imagenet} demonstrates this principle, as increasingly naturalistic datasets 
enabled fair comparisons between diverse model families. This enabled AlexNet~\citep{krizhevsky2012imagenet} to validate deep learning in an era when many researchers doubted its potential.

Thus, we recommend that cognitive scientists establish clear \textit{hold-out} data for both model development and data analysis. This is especially critical for confirming exploratory results~\citep{poldrack2017scanning}.
Combined with standardized evaluation protocols, this strategy enables fair comparison between diverse theoretical frameworks ranging from program synthesis to deep neural networks to reinforcement learning.

\paperarg{R7. Make data accessible with a single line of code}.
\textit{Fully processed} research data should be \textit{programmatically} accessible with a single line of code, requiring no permissions.
While uploading data to platforms like the Open Science Framework (\href{https://osf.io/}{OSF}) is valuable, we additionally recommend platforms like \href{https://huggingface.co/}{Hugging Face} that offer version control, free hosting, authentication-free access, access with one line of code, and uploading data with a few lines of code.
This removes friction from reusing and updating datasets.



\wilka{\paperarg{R8. Code should be easy to ``fork'' and iterate on}.
There should be minimal friction for new researchers to take your existing code and try their ideas on it.
One particularly promising strategy is the ``single file'' philosophy.}
Ideally, key workflows (e.g. model or evaluation definitions)
should be defined in a single file. This has emerged as a powerful force for reproducibility in ML.
While this approach contradicts traditional software engineering principles of modularity and abstraction, it offers distinct advantages for research code: (1) complete understanding of an important component (e.g. a model) requires reading only one file, (2) minimal abstraction layers and dependencies, and (3) easy adaptation of prior methods or their components, by simply \textit{copying} this single file into another codebase. Removing friction from future reuse \andrew{may} also benefit us, \andrew{as more people can build on and cite our work.}


\section{Building from naturalistic experiments to cognitive theories}  \label{sec:theory}

\begin{figure}[htp]
    \begin{center}
        \includegraphics[width=\textwidth]{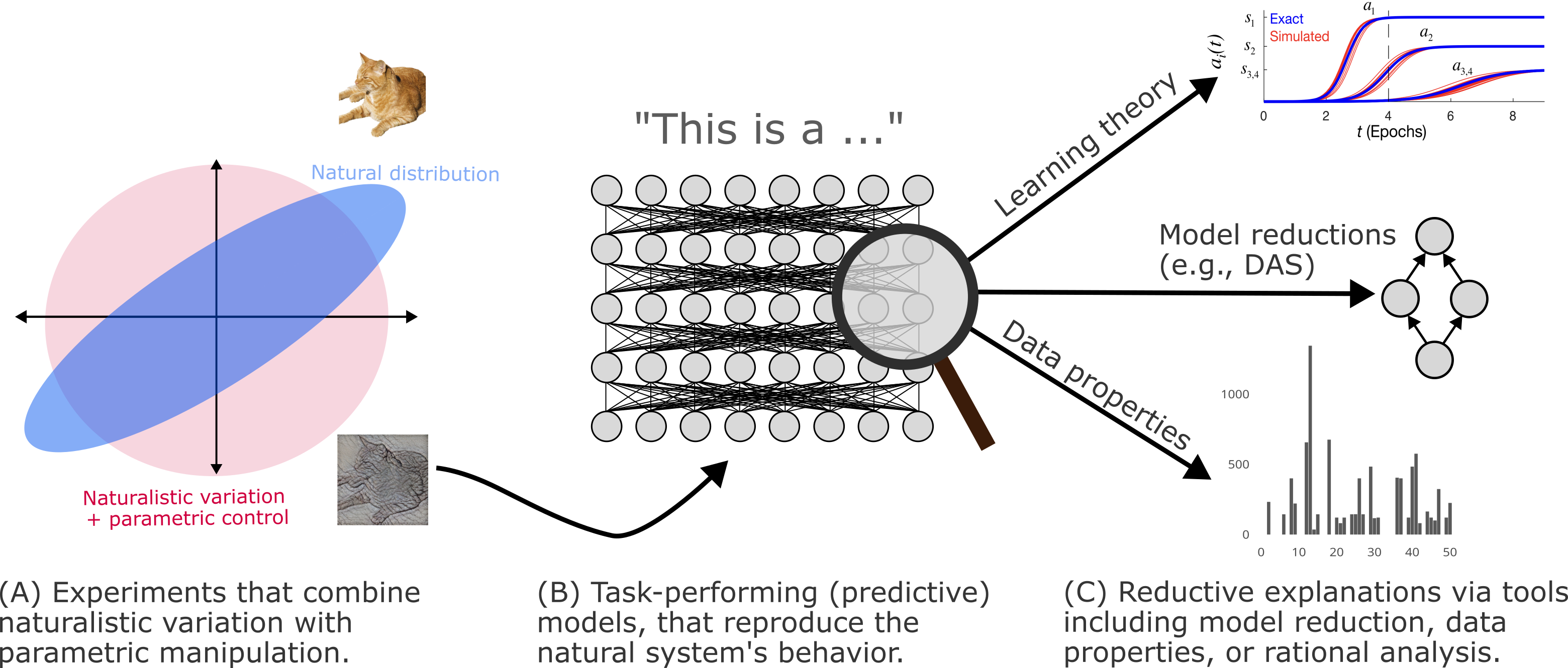} 
        \caption{Overview of how we can develop theories with potentially opaque models and unnatural manipulations of natural data: building task-performing models that can reproduce behavior across naturalistic and unnatural stimuli, while tightly coupling them to reductive explanations. (Panel C top figure is reproduced from \citealt{saxe2019mathematical}; panel A images reproduced from \citet{geirhos2018imagenet}.}
        \label{fig:overview-theory}
    \end{center}
\end{figure}

A primary goal of cognitive science is understanding. 
By instantiating our theories in a model, we are forced to make our theories precise by concretizing the ambiguous details of the mapping from a verbal hypothesis to its implementation~\citep{guest2021computational}.

From this perspective, building computational models that perform the task in as naturalistic a way as possible, across as wide a variety of settings as possible, imposes much stronger constraints on our theories than an abstracted model at a higher level \citep[cf.][]{cao2021explanatorypt2}. These added constraints are important; otherwise, high-level theories that do not directly engage with the details of the implementation can be so unconstrained as to lack explanatory value \citep[cf.][]{erev20151800,jones2011bayesian,rahnev2018suboptimality,andrews2021math}, can be intractable for real problems \citep{van2008tractable}, or can elide important details (see above). 

Yet moving towards naturalistic settings, and the complex models they require, alters the way that we need to approach experimentation and theory building.  Here, we discuss the consequences for experiment design (\S \ref{sec:theory:controlled_experiments}-\ref{sec:theory:incorporating_variation}), the role and interpretation of models (\S\ref{sec:theory:interpreting_models}), and theory building (\S \ref{sec:theory:combining_prediction_with_explanation}).

\subsection{Performing controlled experiments by parametrically manipulating naturalistic data (in unnatural ways)} \label{sec:theory:controlled_experiments}


\begin{figure}[h!t]
\centering
\includegraphics[width=0.6\textwidth]{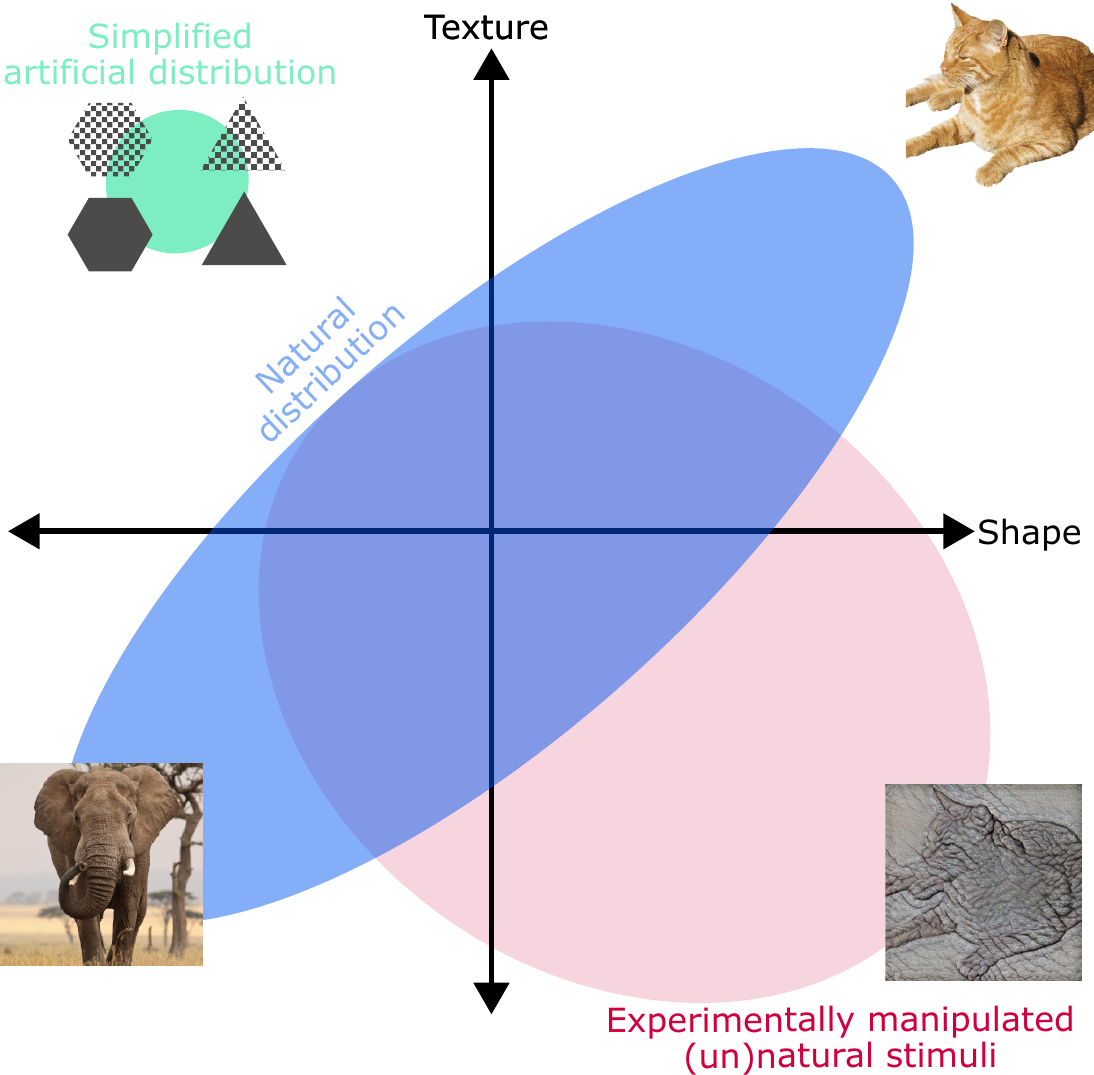}
\caption{Testing hypotheses by parametrically manipulating naturalistic data. In the natural data distribution (blue), shape and texture are highly correlated---cats usually have fur, and elephants usually have wrinkles. This confounding makes it difficult to discern how these variables independently contribute to processing in neural systems or models. One approach to deconfounding these variables is to use a narrow distribution of highly-simplified artificial stimuli (light green) that isolate the variables of interest. However, we advocate for (additionally) creating broader distributions of stimuli that preserve more of the breadth of the natural data distribution, while deconfounding or manipulating features parametrically in order to test hypotheses---as in the unnatural combinations of natural shapes and textures created by \citet{geirhos2018imagenet}. } \label{fig:data_dimensions_controlled_experiments}
\end{figure}

Developing and testing models on truly naturalistic data introduces new issues \andrew{compared to} simple artificial settings. Naturalistic data distributions may be too ``easy''---e.g. they may only rarely include edge-cases that test key capabilities \citep{zhang2023deep}---thus, testing average performance over naturalistic data may disguise important failures. Relatedly, in naturalistic data, features may be confounded, which can prevent readily disentangling the solutions a model uses. For example, models trained on ImageNet \citep{deng2009imagenet} rely on features like texture more than humans do \citep{geirhos2018imagenet}, but 
still perform similarly on ImageNet. These issues limit our ability to achieve deep understanding \andrew{when} experimenting with only the \andrew{naturalistic distribution} \citep{rust2005praise}. 

However, these challenges are surmountable by parametrically manipulating rich, naturalistic data (Fig. \ref{fig:data_dimensions_controlled_experiments})---preserving or enhancing its richness, while still carefully manipulating theoretically-motivated constructs \citep[cf.][]{jain2024computational}. 
For example, \citet{geirhos2018imagenet} engineered datasets that combine natural shapes and textures in unnatural ways. To do so, the authors exploited the contemporary ML technique of iterative style transfer \citep{gatys2016image}---combining features of different images at different spatial scales---thus illustrating how progress in AI models unlocks new experimental approaches. These conflicting shape-texture datasets helped to inspire similar experiments on humans \citep{jagadeesh2022texture,jagadeesh2022human}; surprisingly, those studies found that human visual cortex likewise appears to use textural rather than shape-focused representations. \andrew{Thus,} shape-driven behavior is mediated by downstream readout processes. Thus, manipulating \andrew{naturalistic} features to test models can lead to new insights into brain function.

\andrew{Many} studies similarly systematically manipulate natural stimuli, while preserving as much natural variation as possible---whether introducing new parameters like modality manipulation \citep[e.g. listening vs. reading;][]{deniz2019representation}, inserting challenging syntactic structures in natural stories \citep{futrell2021natural}, or systematically selecting stimuli that drive particular responses while preserving variety \citep[e.g.][]{tuckute2024driving,hosseini2023teasing}. Theory-driven, controlled experimental manipulations can be performed even in more complex, naturalistic settings. 

Another direction for parametrically manipulating naturalistic data comes from using video games to study sequential decision-making and higher-level cognition \citep{allen2024using}.
\andrew{Games combine rich environments with the ability to parametrically manipulate features to reflect naturalistic values, while introducing experimental manipulation. For example }
\andrew{\citet{carvalho2025preemptive} tested human goal preplay using}
Craftax~\citep{matthews2024craftax}---a 2D Minecraft environment \andrew{that} offers parametrically-generated open-world environments containing complex goal hierarchies. \andrew{By sampling complex maps, while constraining possible paths and goal locations, the authors created varied naturalistic experiments that could nevertheless cleanly disentangle different models of goal generalization.}  
\andrew{For researchers who want to build on games,} there is a growing ecosystem of tools for building complex naturalistic experimental paradigms, and evaluating humans and models with them\andrew{--- such as NiceWebRL\footnote{\url{https://github.com/KempnerInstitute/nicewebrl}}, which enables web-browser human experiments using ML environments.}

Moreover, expanding the naturalism of our experiments \emph{enhances} our ability to develop generalizable theories, as we argue next.

\subsection{Incorporating naturalistic variation to enhance generalizability}   \label{sec:theory:incorporating_variation}

By testing hypotheses in settings where other aspects of the task and data generating process are sampled broadly from the naturalistic distribution, we address some of the generalizability challenges that arise from testing on a much narrower distribution than the written hypothesis implies \citep[cf.][]{yarkoni2022generalizability}; 
``design[ing] with variability in mind'' \citep{yarkoni2022generalizability} increases the likelihood of identifying robust effects that will generalize \citep[cf.][]{baribault2018metastudies}. In the language of regression, varying as many other data generating factors as possible brings us closer to estimating the ``main effect'' of an experimentally-manipulated construct, rather than estimating the ``simple effect'' of the manipulation in the single setting we have tested. 

For example, it is a long-standing principle of experimental design to test a hypothesis with multiple stimuli, and statistically quantify the stimulus effect to estimate generalizability \citep[e.g.][]{clark1973language}. However, many other experiment features---even exact task formulations such as multi-arm bandit tasks---are often preserved within a paper, or even across an entire literature. Just like testing a hypothesis with a single stimulus, this limited variation poses challenges for generalizability \citep{eckstein2022interpretation}---especially because the way that researchers operationalize a theory can substantially affect their conclusions \citep[e.g.][]{holzmeister2024heterogeneity,schweinsberg2021same,strickland2012experimenter}. While naturalistic variation cannot resolve this issue in full, we believe that incorporating a broader range of the settings, stimuli, and tasks within which we expect a theory to hold, will help us to develop more generalizable theories. 

Incorporating variation can also help us to revise the constructs underlying our theories. For example, \citet{eisenberg2019uncovering} study the putative construct of self-regulation by simultaneously testing many existing self-regulation measures. The authors find that these measures do not form a unitary construct. Rather, task- and survey-based measures each form distinct, multi-factorial constructs, with individual tasks loading primarily onto a subset of factors. That is, two different studies that each use only a single measure for self-regulation may be measuring entirely different constructs. By contrast, by substantially varying the tasks we use to measure a theoretical component like self-regulation, we can more fully disentangle the underlying constructs---and thereby improve the extent to which our theories generalize. \andrew{Likewise, \citet{peterson2021using} collect a large dataset of risky choice judgments, and use it to explore how human choices interpolate between different prior decision-making theories depending on the problem context.}

\subsection{Interpreting complex models to yield explanations} \label{sec:theory:interpreting_models}

\andrew{How can complex experiments in naturalistic settings, and complex models that can predict behavior across them, }support explaining the cognitive phenomena of interest? Here, we illustrate some paths to deriving explanations from complex models.\footnote{N.B., not all complex systems admit abstract explanations. For example, where natural intelligence incorporates chaotic \citep{freeman1995chaos} or critical \citep{obyrne2022critical} dynamics, they may alter the explanations that we can seek \citep[cf.][]{kellert1993wake}. A benefit of seeking predictive models as a route to explanation is that the models may be useful even if simple explanations do not follow.}

\paperarg{Data properties as explanation \& rational analysis.} One way to explain model behavior is as a consequence of the properties of the data from which it learns. 
This idea has a rich history in cognitive science and AI \citep{mcclelland1991nature,elman1996rethinking,rogers2004semantic}, in which researchers attempted to identify minimal properties of naturalistic data that give rise to some behavior(s).
\andrew{More recently, these techniques have been applied to understanding AI models.} For example, \citet{chan2022data} explore how the bursty, long-tailed nature of natural data can give rise to in-context learning, and \citet{prystawski2024think} examine how local dependencies can yield certain kinds of sequential reasoning. \andrew{Likewise, \citet{hermann2020origins} demonstrate how properties of data created by augmentation strategies can drive the texture bias investigated with parametric manipulations by \citet{geirhos2018imagenet}, as discussed above.} These works illustrate how rich properties of natural data can be distilled down to minimal elements that drive a behavior \andrew{or computational strategy}---and can thus offer a candidate explanation for the origin of those capabilities \andrew{or biases}. Moreover, explanations of behavior via properties of the data \andrew{point towards} \emph{normative} explanations of that behavior as a rational solution to constraints---precisely the perspective taken by \emph{rational analysis} \citep{anderson1991human,lieder2020resource}.

\andrew{For example, \citep{lampinen2024language} compare language models and humans on various logical reasoning tasks---from natural language inference to syllogism validity judgments \citep{evans1983conflict} and the Wason selection task \citep{wason1968reasoning}---instantiated in both synthetic and naturalistic settings. The authors find that models and humans show common patterns of content-entangled reasoning and graded uncertainty. By studying these common patterns, the authors argue that prior interpretations of particular phenomena (e.g., social explanations of the Wason selection task; \citealp{cosmides1989logic}) were overfit to the experiments explained. Instead, the authors argue for a unifying (resource) rational account of these superficially-different biases as stemming from the underlying statistical properties of natural data for both humans and models: content effects arise from optimizing prediction over natural data where certain conclusions tend to hold much more often than others. This example shows how applying rational analysis across naturalistic settings can generate unifying accounts of seemingly-disparate phenomena.}

\paperarg{Mechanistic explanation.} Understanding that a behavior arises from data properties does not explain \andrew{its implementation}. Fortunately, the internal workings of computational models can generally be inspected and intervened upon. For example, many works have analyzed the mechanisms that implement language model behaviors, either in simplified settings \citep[e.g.][]{ nanda2023progress,zhong2024clock}, or in large models trained on natural data \citep{geiger2021causal,wu2024interpretability}. A particularly promising approach is illustrated by Distributed Alignment Search \citep{geiger2024finding}, which maps postulated abstract causal models onto low-level features of a complex model that play corresponding causal roles---thus enabling linking between abstract models or computational hypotheses to evidence about concrete implementational mechanisms.

Moreover, mechanistic interventions can even be combined with data studies, to examine the causal role of different mechanisms over learning \citep{singh2024needs}. Appropriate regularization of models can also \andrew{yield more interpretable representations}---e.g. \citet{miller2024cognitive} train networks with regularized representations to predict animal behavior, and used them to identify non-optimal features influencing the decisions---thus allowing bottom-up discovery of possible implementations. Thus, various methods for mechanistic study of models can offer powerful routes to experimentally determining sufficient implementations of cognitive processes.

\paperarg{Formal theories.} From data properties or mechanisms, we can progress to formal theories. \citet{saxe2019mathematical} offers an illustrative example. Prior works observed empirically that pseudo-naturalistic data elicit various aspects of human-like semantic development and representation in neural networks \citep{rogers2004semantic}, such as progressive differentiation. However, these works had not explained \emph{why} these parallels emerge. \citet{saxe2019mathematical} bridged this gap by formally deriving how the data properties combine with gradient-based learning to yield these cognitive phenomena. This illustrates how preliminary explanations in terms of data can be developed into more rigorous analytic theories. Thus, building naturalistic models can be a stepping stone toward more rigorous, normative theories of intelligence as a rational solution to a particular problem setting---much like prior work in cognitive science 
\citep[e.g.][]{anderson1991human,oaksford2009precis}.

\andrew{
\paperarg{Bayesian or other reductive cognitive models.} Another analytic approach is to interpret complex models using the very tools of cognitive modeling, for example by building computational-level Bayesian models \citep{griffiths2024bayes,ku2025using}. Indeed, various recent works have taken the approach of reducing from more complex task-performing models to simpler models based on prior cognitive work. For example, \citet[][discussed above]{peterson2021using} initially fit an unconstrained context-dependent neural model of the behavior, but reduced it to a similarly-predictive model that effectively does context-dependent interpolation between prior decision-making theories---thus providing a unifying but relatively interpretable account of a broader range of phenomena \citep[cf.][]{fintz2022using}. As an example of the Bayesian approach, \citet{zhu2024incoherent} show how language models produce some human-like \citep{zhu2020bayesian} patterns and biases in their probabilistic judgments. The authors link these patterns of behavior to a computational-level Bayesian model---and formally relate that model to the probabilistic inference performed by the language models. These examples, like those in the prior section, show how it is possible to build explicit links from complex task-performing models to simpler, more classic computational cognitive models.}

\subsection{Theories that \andrew{couple} task-performing models with reductive explanations.} \label{sec:theory:combining_prediction_with_explanation} 
Above, we have argued for the value of embracing the richness of naturalistic tasks, even if it leads us to build more complex task-performing models that are hard to interpret, but that perhaps make more accurate predictions of behavior. In the previous section, we have outlined some of the possibilities (and challenges) of deriving understanding from these models. So what, ultimately, should we seek?

We argue that naturalistic computational cognitive science should seek theories of cognitive phenomena that consist of two components:
\begin{enumerate}
\item Task-performing (predictive) models that reproduce the phenomena across the same range of naturalistic stimuli and paradigms as the human/animal subjects.
\item Reductions of these task-performing models to simpler mechanisms, properties, and theories of \emph{why} these models reproduce the phenomena.
\end{enumerate}
These two components serve different purposes. The first \andrew{component---\textbf{a task-performing model}---}helps to demonstrate that the \andrew{posited theory offers a real candidate account} of the \andrew{full range of natural} phenomena in question, rather than \andrew{relying on approaches} that simply could not scale to the real problem or that are overfit to a particular instantiation of the task. \andrew{Conversely, this component increases the probability that our model accounts for all the possible constraints the real stimuli may impose on human learning and task behavior.} \andrew{As outlined above,} this component is necessary because otherwise the problem space is under-constrained. Moreover, it provides a test-bed for experiments that would be impossible in reality, which is useful both scientifically and practically. \andrew{From a scientific perspective, building task-performing models enables testing the model on other phenomena within the same task space, which can enable more unifying accounts of how phenomena relate. It also enables exploring hypotheses that would be impossible to test in reality, such as the effect of removing a particular syntactic construction from a learner's linguistic experience. Practically, being able to simulate responses to real stimuli is also useful for the many applications of cognitive science to practical problems in domains such as education.}

The second component\andrew{---\textbf{reductive explanations}---}allows linking these \andrew{complex predictive models} to the explanatory understanding that cognitive scientists typically seek, and that may help us to generalize our insights beyond current paradigms. \andrew{As outlined above, this component can build on the explanatory approaches of prior Bayesian, connectionist, analytic, or normative theories in cognitive science.} This \andrew{part of our} perspective aligns with past arguments on how deep learning can contribute to understanding in cognitive neuroscience \citep{saxe2021if,kanwisher2023using,cao2021explanatorypt2,doerig2023neuroconnectionist}. However, we emphasize that \andrew{synthesizing} these two components makes explicit\andrew{, testable} links between the actual problem the system solves and the theorized constructs underlying that solution. It also ensures that our theories make testable predictions beyond the narrow settings of a particular task paradigm.

\section{Discussion}\label{sec:discussion}

In this paper, we have tried to outline a direction of research that we call ``naturalistic computational cognitive science,'' that aims to build generalizable models \andrew{and theories} of cognition that scale to naturalistic tasks, while still \andrew{affording} explanatory understanding. This perspective synthesizes a growing literature on the importance of naturalistic experiments and generalizable models, and grows out of a long-standing focus in cognitive science on explaining a broad range of phenomena. 
In this section, we outline these connections, and highlight the future directions suggested by our perspective.

\paperarg{The quest for building general models of natural intelligence.} 
\andrew{Many cognitive scientists have} sought frameworks with the generality to explain a wide scope of cognitive phenomena. Researchers in cognitive architecture \citep[e.g.][]{ritter2019act,laird2019soar}, connectionist \citep[e.g.][]{mcclelland1986appeal,mcclelland2003developing}, and Bayesian \citep[e.g.][]{tenenbaum2001generalization,griffiths2010probabilistic,tenenbaum2011grow}, paradigms have tried to identify underlying principles or mechanisms that explain many phenomena. However, while these modeling frameworks themselves are general, typical \emph{instantiations} of the frameworks have built specialized models focused on individual tasks. Indeed, one core critique of many modeling paradigms is their failure to accurately capture patterns of human generalization \citep[e.g.][]{fodor1975language,mcclelland2002rules,lake2017building}.

In pursuit of generality, some recent studies have built upon foundation models from AI that can perform many tasks. Some works have applied cognitive paradigms to study the behaviors of these models \citep[e.g.][]{binz2023using,lampinen2024language,buschoff2025visual}. Others have finetuned these models using cognitive data, to create generalizable models that can predict behavior on new experimental paradigms in areas like vision \citep[e.g.][]{muttenthaler2024improving,muttenthaler2024aligning,fu2024dreamsim}, or the broader space of cognitive tasks that can be presented in language \citep{binz2024turning,binz2024centaurfoundationmodelhuman}.

We see the arguments that we have laid out in this paper as being broadly compatible with many of these approaches and frameworks. However, we offer a stronger emphasis on the role of theory \citep[cf.][]{frank2025cognitive}, and the virtuous cycle between expanding the scope and naturalism of our experimental designs, and expanding the generalizability of our models and theories.

\paperarg{Towards experiments that involve naturalistic behavior}
Our perspective aligns with various works that have likewise highlighted the value of considering complex naturalistic behaviors and the environments in which they occur. The strands of cognitive research focused on embodiment and enactivism have advocated that the environment, and organism-environment interactions, form an important component of cognition \citep[e.g.][]{clark1998being,varela2017embodied}. More generally, naturalistic experiments are frequently advocated in neuroscience \citep[e.g.][]{mobbs2021promises,cisek2024toward}, where even prominent visual processing signals can fundamentally differ in active, naturalistic paradigms \citep[e.g.][]{amme2024saccade}. Similarly, \citet{gao2015simplicity,gao2017theory} highlight that more complex behavioral paradigms are needed to unlock the full space of neural dynamics, and \citet{nastase2020keep} argue that isolating only a few variables in our experiments impairs our ability to understand the full scope of natural behavior. Likewise, \citet{wise2023naturalistic} advocate for the need to explore more naturalistic environments and behaviors in reinforcement learning, to grapple with the complexity of behavior in environments closer to the real world. We concur with these works, and emphasize that advances in technology both enable richer experimental paradigms and provide the models that can perform them. We differ from many of these works in more strongly emphasizing the role of models that generalize across task paradigms.

\paperarg{Linking between learning models and the brain}
A large recent literature has explored surprising alignments between the representations learned by task-optimized deep learning models and neural representations \citep{khaligh2014deep,yamins2014performance,yamins2016using,schrimpf2021neural,sucholutsky2023getting}. These findings have stimulated substantial interest and debate, ranging from critical arguments that failures to capture particular phenomena doom these models \citep[e.g.][]{bowers2023deep}, to suggestions that deep learning models upend the theoretical assumptions of the field \citep{perconti2020deep,hasson2020direct}.
Other authors have written frameworks that attempt to integrate deep learning models within the existing scientific paradigms \citep{doerig2023neuroconnectionist}.
In this vein, several papers have proposed frameworks for understanding the role of deep learning in cognitive neuroscience \citep[e.g.,][]{richards2019deep,storrs2019deep,cichy2019deep}---and how these models can contribute to deriving explanatory and theoretical understanding \citep{cao2024explanatorypt1,cao2021explanatorypt2,saxe2021if}. In this context, \citet{feather2025brain} argue for the importance of benchmarks that test alignment of both behavior and representations---and highlight how dataset limitations like lack of stimulus variability can limit our ability to discriminate between models. As above, our perspective aligns with many of these works, but we place greater emphasis on the complementary role of increasing naturalism of experimental paradigms, and learning from the engineering paradigms of AI, rather than simply adopting its models as artifacts.

\paperarg{Underdetermination in science} 
\andrew{Our arguments touch on broader issues in the philosophy of science. In particular, our scientific theories are \emph{always} underdetermined by the finite experiments that we have performed \citep{stanford2023underdetermination}---our conclusions about what an experiment (dis)proves hinge upon a collection of background assumptions, e.g. about the methods involved, the systems being experimented upon, etc. We see this paper as highlighting several common background assumptions underlying computational cognitive science that we believe are suspect and can be detrimental to building effective theories---in particular, the assumptions that simplified experimental settings can allow building theories of a system's behavior that will generalize consistently to more naturalistic ones, and that modeling experiments over simplified stimuli can accurately characterize the computational challenges of learning and generalizing over naturalistic stimuli. In the earlier sections of the paper, we provided illustrative examples of how these assumptions can break down; in the later sections we advocated for approaches that can avoid making these assumptions. While our proposed methods (like any empirical approach) still rely on other background assumptions---e.g., about the fidelity of the links we can make between naturalistic models and the system being modeled---we hope that by avoiding background assumptions that have been empirically found to be problematic, we can make some progress towards more robust and general cognitive theories.}

\subsection{Looking forward: challenges \& opportunities for naturalistic computational cognitive science}
While we have advocated for the importance of naturalism, there are a number of practical and conceptual challenges to achieving these goals. Many of these challenges provide important opportunities for future research to address.

First, as we noted above, identifying the parameters of naturalistic variation that may interact with a particular theoretical construct of interest can be challenging. When we form theories---for object perception, for theory of mind, for how people plan---which ``parameters'' of the experimental paradigm should we make more naturalistic? In the trolley problem, for example, is it the number of people that must be considered? Is it the way in which people might die? Is it the length of time to decide?  While researchers can draw on domain expertise to identify potentially-relevant parameters, there may be unexpected influences from other factors that are not typically manipulated--e.g. the differences between moral thoughts and moral actions when shifting from vignettes to virtual reality \citep{francis2017simulating}. Thus, one important direction for naturalistic research may be to identify more general overarching naturalistic factors---such as  perception, embodiment, or action---that tend to influence phenomena across many domains. Ultimately, identifying such overarching factors may help us to identify more general unifying principles of cognition.

Another challenge and opportunity is precisely specifying timescales of learning.
Current learning-based theories often conflate evolution, development, and short-term learning into a single process of ``optimizing'' a model for the task of interest. This is motivated from a rational analysis perspective, where the artifact of interest is the outcome of the optimization process, which serves as a theory for behavior. While this has led to fruitful understanding, 
it leaves open the question of what exactly is learned during each stage. 
More complete ``learning''-based accounts would specify precisely what an ``evolutionary'' data distribution would be, and the model behaviors (evolved inductive biases) that are expected from it; a succession of ``developmental'' data distributions and the model behaviors that would result from them; and finally predict both the behaviors that people begin tasks with and how they change with task exposure.
Fortunately, work in this direction has already begun.
For example, researchers have begun to explicitly map different timescales of learning to different levels of optimization in a meta-learning process \citep{wang2021meta}---and indeed, some works have highlighted how meta-learning can give rise to human-like inductive biases, e.g. for linguistic \citep{mccoy2020universal,mccoy2025modeling} or compositional \citep{lake2023human} generalizations. Thus, a key opportunity for naturalistic computational cognitive science is to develop learning-based account of cognition that begin to specify how learning from different data-distributions within different segments of learning can produce the changing cognitive phenomena we observe throughout ontogeny.

Finally, the approaches we have advocated for introduce practical challenges. The types of modeling and experimental paradigms we discuss here often either push towards or pull from the frontiers of contemporary computer science research. This can require both greater technical expertise and computational resources. While we acknowledge this challenge, we see it as an opportunity for larger-scale collaborative efforts---which are increasingly adopted in computer science research on training large-scale models \citep[e.g.][]{le2023bloom}. By pooling expertise and resources, collaborations make progress at larger scales than individual research groups could. Likewise, as discussed above, collaborations on shared benchmarks \citep[e.g.][]{schrimpf2020integrative} can allow broader and deeper assessments than any individual researcher could. Furthermore, the growing overlap between cognitive modeling and artificial intelligence brings opportunities for mutually-beneficial collaborations between the two fields. Cognitive scientists have unique expertise in developing evaluations for scientifically inferring what cognitive algorithms may underly behavioral phenomena; on the other hand AI researchers have unique expertise in training models that can perform tasks of interest in naturalistic settings. We hope that this paper will help to encourage deeper collaborations between these fields on their shared questions about the nature and origins of intelligence. 

\subsection{Conclusions}
We have outlined a perspective that advocates for studying cognition through a broad spectrum of naturalistic experimental paradigms, building generalizable models of cognition that can perform naturalistic tasks, and deriving explanatory theories from these models and their simplifications. We have supported this perspective with examples illustrating the value of naturalistic settings and generalizable models, and concrete guidance on the practicalities of building generalizable models and using them to shape our theories. Many individual aspects of our argument align with prior works, but we hope that there is value in synthesizing these perspectives and highlighting their synergy. We hope that this perspective will inspire computational cognitive scientists to continue embracing richer naturalistic experimental paradigms and models.

\clearpage
\subsection*{Acknowledgements}
We first thank Sam Gershman for his extensive guidance and feedback on this article and on how to integrate cognitive science and machine learning methodologies.
We also thank George Alvarez, Maria Eckstein, Tyler Bonnen, Jay McClelland, Mike Mozer, Rachel Calcott, Nick Blauch, Aran Nayebi, Ruben Cohen, Greta Tuckute, Fenil Doshi, and Shuze Liu for thoughtful comments and discussions. Finally, we thank Felix Hill for many inspirational conversations on generalization, cognition, and more.

\subsection*{Funding statement}
This work was supported
by a gift from the Chan Zuckerberg Initiative Foundation
to establish the Kempner Institute for the Study of Natural
and Artificial Intelligence. 

\subsection*{Conflicts of interest statement}

The second author is employed by Anthropic.

\bibliographystyle{apalike}
\bibliography{bib}

\end{document}